\documentclass[aps,twocolumn,showpacs,superscriptaddress]{revtex4}

\usepackage[T1]{fontenc}
\usepackage[utf8]{inputenc}
\usepackage{lmodern}
\usepackage{microtype}

\usepackage{mathtools}
\usepackage{amssymb}
\usepackage{bm}
\usepackage{dcolumn}

\def\EE{{\cal E}}
\def\LL{{\cal L}}

\newcommand{\beq}{\begin{equation}}
\newcommand{\eeq}{\end{equation}}
\newcommand{\bea}{\begin{eqnarray}}
\newcommand{\eea}{\end{eqnarray}}

\usepackage{graphicx}
\usepackage[dvipsnames]{xcolor} 
\usepackage[normalem]{ulem}     
\usepackage{mathrsfs}
\usepackage{scalerel}
\usepackage{tikz}
\usetikzlibrary{svg.path}
\definecolor{orcidlogocol}{HTML}{A6CE39}
\tikzset{
  orcidlogo/.pic={
    \fill[orcidlogocol] svg{M256,128c0,70.7-57.3,128-128,128C57.3,256,0,198.7,0,128C0,57.3,57.3,0,128,0C198.7,0,256,57.3,256,128z};
    \fill[white] svg{M86.3,186.2H70.9V79.1h15.4v48.4V186.2z}
svg{M108.9,79.1h41.6c39.6,0,57,28.3,57,53.6c0,27.5-21.5,53.6-56.8,53.6h-41.8V79.1z M124.3,172.4h24.5c34.9,0,42.9-26.5,42.9-39.7c0-21.5-13.7-39.7-43.7-39.7h-23.7V172.4z}
svg{M88.7,56.8c0,5.5-4.5,10.1-10.1,10.1c-5.6,0-10.1-4.6-10.1-10.1c0-5.6,4.5-10.1,10.1-10.1C84.2,46.7,88.7,51.3,88.7,56.8z};
  }
}
\usepackage{orcidlink}
\usepackage{hyperref}
\hypersetup{
  colorlinks=true,
  linkcolor=blue,
  citecolor=cyan,
  urlcolor=MidnightBlue,
  breaklinks=true
}
\usepackage{doi} 
\usepackage{float}
\newcommand\orcidicon[1]{\href{https://orcid.org/#1}{\mbox{\scalerel*{
\begin{tikzpicture}[yscale=-1,transform shape]
\pic{orcidlogo};
\end{tikzpicture}}{|}}}}
\begin{document}
\title{Charged Simpson-Visser-AdS spacetime: QPOs and observational constraints}
\author{Saeed Ullah Khan \orcidlink{0000-0001-9468-0104}}
\email{saeedkhan.u@gmail.com}
\affiliation{School of Mathematics and Statistics, Fuzhou University, Fuzhou 350108, Fujian, China}
\author{Javlon Rayimbaev\orcidlink{0000-0001-9293-1838}}
\email{javlonrayimbaev6@gmail.com}
\affiliation{Institute of Theoretical Physics, National University of Uzbekistan, Tashkent 100174, Uzbekistan}
\affiliation{Kimyo International University in Tashkent, Shota Rustaveli Street 156, Tashkent 100121, Uzbekistan}
\affiliation{Tashkent State Technical University, Tashkent 100095, Uzbekistan}
\author{Muhammad Zahid}
\email{zahid.m0011@gmail.com}
\affiliation{Department of Physics, Shanghai University, Shanghai, 200444, China}

\author{Tolaniddin Nurmukhamedov\orcidlink{0000-0002-2507-3674}}\email{ntolaniddin@mail.ru} 
\affiliation{Tashkent State Transport University, Tashkent 100067, Uzbekistan}

\author{Abat Muratov\orcidlink{0000-0001-6440-9460}}
\email{acmuratov@mail.ru}
\affiliation{Karakalpak State University, Sh. Abdirov 1, Nukus 230112, Uzbekistan}

\author{Weiwei Wang}\email{pde_fzu@163.com}
\affiliation{School of Mathematics and Statistics, Fuzhou University, Fuzhou 350108, Fujian, China}
\date{\today}

\begin{abstract}
In this article, we examine particle dynamics and Quasi-periodic oscillations within a charged Simpson-Visser-Anti-de Sitter (AdS) spacetime. We first analyze the behavior of the metric and black hole horizons under the competing impact of electric charge $Q$, regularization parameter $b$, and AdS radius $l$. The roots of the metric function reveal transitions between three distinct topological classes: non-extremal configurations featuring distinct inner and outer horizons, extremal limits with a single degenerate horizon, and strictly positive, horizonless wormhole geometries. By evaluating the effective potential, we determine how this modified geometry shifts the Innermost Stable Circular Orbit (ISCO) and alters the angular momentum and energy of test particles. Moreover, we derive the Keplerian, radial, and vertical frequencies to model quasi-periodic oscillations (QPOs). We show that the interplay between the regularization parameter $b$ and the AdS radius $l$ significantly modifies twin-peak QPOs. In particular, $b$ drives fundamental QPO frequencies higher than those predicted by standard general relativistic backgrounds. Twin-peak high-frequency QPO data from two black-hole binaries disfavor the Schwarzschild limit at $\sim\!4\sigma$, but constrain only a degenerate combination of charge and AdS curvature, leaving the regularization parameter entirely unconstrained by the data.
\end{abstract}
\maketitle

\sloppy
\section{Introduction}\label{Sec:introduction}
Theories of gravity provide the fundamental framework for understanding the origin, evolution, and large-scale dynamics of the universe, as well as the behavior of compact relativistic objects including neutron stars and black holes.
In 1915, Einstein first introduced the general theory of relativity (GR) to incorporate non-inertial reference frames into special relativity, describing gravity as a geometric manifestation of spacetime curvature. GR has been rigorously verified as the standard theory of gravity across both weak-field regimes with the help of solar system tests and weak gravitational lensing~\cite{Amendola2008JCAP...04..013A}, and strong-field regimes, as demonstrated by gravitational wave detections~\cite{Pitkin2011LRR.14..5P,Liu2024PhRvD.109h4074L} and direct horizon-scale imaging of the supermassive black hole shadows Sgr A*~\cite{EHT2022ApJ...930L..12E,EHT2022ApJ...930L..17E} and M87*~\cite{EHT2019ApJ...875L...1E,2023MNRAS.523..375S,EHT2023ApJ...957L..20E}.

Despite these successes, GR has theoretical limitations, most notably spacetime singularities characterized by divergent curvature and energy density, as well as its fundamental incompatibility with quantum field theory. These open problems motivate modified gravity models and alternative gravitational theories. However, observational parameter degeneracies often make it difficult to distinguish between distinct theoretical formulations based on astrophysical signatures alone~\cite{2021JCAP...04..082M,2023MNRAS.523..375S}. To address this, researchers have introduced model-independent parameterizations of field-equation solutions, particularly for general spherically and axially symmetric spacetimes~\cite{Rezzolla2014PhRvD,Konoplya2020PhRvD.101l4004K}.

In~\cite{Simpson2019JCAP...02..042S} Simpson and Visser suggested an elegant approach for avoiding the center singularity. This spherically symmetric meta-geometry, including a generalized rotational variant~\cite{2021JCAP...04..082M}, describes either a regular black hole or a traversable wormhole, depending on the regularization length-scale parameter $b$. This parameter may signify the influence of quantum gravity or other yet unidentified phenomena. In the instance of a wormhole, a conduit between two points in the universe (or across distinct worlds) resembles a conventional Morris-Thorne wormhole. Recall that a traversable wormhole produced this way often requires a unique exotic substance that violates the weak-energy criterion by having negative energy density and tension in its center region. However, wormholes have also been proposed without exotic matter, such as the Einstein-Dirac-Maxwell wormhole \cite{2021PhRvL.126j1102B,2022PhRvL.128i1104K,2022EPJC...82..533B,2023EPJC...83..854V}. In Ref.~\cite{Ahmed2026EPJC...86..658A}, the Simpson-Visser regularization approach has been extended to Anti-de Sitter charged black holes to study their geodesic structure and thermodynamic behavior. Recently, Khan et al.~\cite{Khan2026ChJPh.102..711K}, extended the Simpson–Visser back hole to its rotating counter part in modified gravity theory.

Any alterations or modifications to GR must undergo testing. Analyzing the behavior of the test particle represents the most efficient and practical approach of assessment. Examining the dynamics of both massive and massless particles inside any metric theory of gravitation might provide significant insights into the physical characteristics of the solution~\cite{Chandrasekhar1983mtbh.book.C,Bambi2017bhlt.book..B}. X-ray data from astrophysical sources facilitate the study of compact-object solutions~\cite{Wilkins2012MNRAS}.
The dynamics of neutral and charged test particles near black holes is essential in astrophysics and has been extensively studied to probe background geometry by various authors ~\cite{Aliev2002MNRAS.336..241A,Vrba2020PhRvD.101l4039V,Zahid2022EPJC...82..494Z,2022ChJPh..78..141K,khan2024circular}.

The Rossi X-ray Timing Explorer (RXTE) research project is a valuable and useful resource for studying X-ray binaries. The effort resulted in many black hole transient observations~\cite{1993ARA&A..31...93V,2012MNRAS.426.1701B}. 
This subject is exciting because it allows deeper exploration of fundamental physics. In binary systems with stellar-mass black holes, QPOs in X-ray flux curves are still being studied intensively. These oscillations are regarded as a successful test of strong-gravity theory~\cite {2022EPJC...82.1110R, 2022CQGra..39g5021R, 2023EPJC...83..730Q, Rayimbaev2023EPJC...83..572R, 2022PDU....3500930R, 2023Galax..11...95R}. These changes have frequencies that closely match the black hole's activity and are inversely proportional to its mass. 
Several varieties of QPOs were found using measured frequencies ranging from a few MHz to $0.5$ kHz. They often match high-frequency (HF) QPOs with maximum frequencies of 500 Hz and low-frequency (LF) QPOs with maximum frequencies of $30$ Hz.

The RXTE research project discovered various complex, irregular structures, including QPOs above 40 Hz~\cite{2012MNRAS.426.1701B}. HF-QPOs give insights on the spin and masses of core entities~\cite{2001A&A...374L..19A} and the radii and masses of neutron stars~\cite{1998ApJ...499L..37M,1998ApJ...509L..37K}. The HF-QPO frequencies correspond to the epicyclic frequencies of particle motion in the innermost stable circular orbit (ISCO). They have therefore received substantial theoretical attention from numerous researchers~\cite{Jumaniyozov2024EPJC964y,2026EPJC...86..510S}. Most recently, Khan et al.~\cite{Khan2026EPJC...86..597K} studied the QPOs in a Kerr-like rotating Simpson–Visser black hole in MOG. Although researchers have studied QPOs in Simpson-Visser and modified-gravity black holes~\cite{Rayimbaev2021Galax.9.75R,2025EPJC...85.1029N,Stuchlik2021JCAP...11..059S,2025PhRvD.112l4018D,Li2026arXiv260214458L}, our manuscript presents an interesting physical interaction not present in single-parameter modifications. In the Simpson-Visser framework, the regularization parameter $ b$ determines the upper and lower orbital frequencies ($\nu_u, \nu_L$) alone; here, we also consider the effects of the electromagnetic charge $Q$ and the AdS radius $l$.

The manuscript is organized as follows: in section~\ref{Sec:Charged-SV-AdS}, we briefly introduce and examine the charged Simpson-Visser-AdS spacetime geometry. In section~\ref{sec3: circular motion}, we explore the circular motion: the effective potential, the specific angular momentum and energy obtained via critical orbits, and the ISCO of a charged Simpson-Visser-AdS black hole. Moreover, we will study the fundamental frequencies and QPOs in section~\ref{sec4:Fund_frequencies and QPOs} of our manuscript. 
Finally, in the last section~\ref{sec:conclusion}, we conclude our findings with concluding remarks.
\section{Charged Simpson-Visser-Ads spacetime}\label{Sec:Charged-SV-AdS}
In this section, we examine the charged Simpson-Visser-AdS black holes. Following the Simpson and Visser regularization approach, one can avoid the black hole singularity to get a regular black hole geometry everywhere. Therefore, to obtain a regular Simpson-Visser spacetime, we use the Simpson-Visser regularization technique via the transformation $r^2\to r^2+b^2$~\cite{2021JCAP...07..036F}. This extends the doamin of $r$ from $r \in (0, +\infty)$ to $r \in (-\infty, +\infty)$ for a non-zero value of $b$. Here, $b$ is the regularization parameter, also known as the Simpson-Visser regularization parameter. As a result, with the help of the Simpson-Visser regularization approach, the line element in the coordinates $x^\mu = (t, r, \theta, \phi)$ delineates the geometry around a static Simpson-Visser spacetime and reads
\begin{equation}\label{metric}
ds^2=-f(r)dt^2+\frac{1}{f(r)}dr^2+(r^2+b^2)d\Omega^2\ , 
\end{equation}
with  $d\Omega^2 = d\theta^2 + \sin^2\theta d\phi^2$, and
\begin{equation}\label{metric1}
f(r)=1 - \frac{2M}{\sqrt{r^2 + b^2}}.
\end{equation}
In this equation, $M$ is the ADM mass, and $b$ is the regularization parameter. The event horizon of the above Simpson-Visser black holes may be determined by solving $f(r)=0$, as follows: 
\begin{equation}
{r_h}= \pm \sqrt{(2M)^2-b^2}.
\end{equation}
We can see that $b< 2M$ represents a regular black hole, while $b\geq 2M$ yields a wormhole.

The electromagnetic four-potential $A_{\mu} = - Q \delta_{\mu}^t/{r} $ describes the Reissner––Nordström––AdS (RN-AdS) black hole, a generalization of the Schwarzschild-- AdS solution with a non-zero electric charge $Q$. Henceforth, the corresponding spacetime geometry of RN-AdS black holes can be defined with the help of the following lapse function:
\begin{eqnarray}\label{function-RN-AdS}
 f(r)=1-\frac{2M}{r}+\frac{Q^2}{r^2}+\frac{r^2}{l^2}.   
\end{eqnarray}
Here, $l$ denotes the AdS radius, related to the cosmological constant by $\Lambda =- {3}/{l^2}$. 
To obtain the corresponding modified lapse function, we again use the Simpson-Visser regularization approach via the transformation $r^2\to r^2+b^2$~\cite{2021JCAP...07..036F}, referred to as the charged Simpson-Visser-AdS spacetime,
\begin{equation}\label{function_f}
f(r)=1 - \frac{2M}{\sqrt{r^2 + b^2}} +\frac{Q^2}{r^2+b^2} + \frac{r^2 + b^2}{l^2}.
\end{equation}
The electromagnetic four-potential for the above regular spacetime geometry is modified to $A_{\mu} = - Q \delta_{\mu}^t/\sqrt{r^2+b^2}$.

The above metric in Eq.~\eqref{metric}, reduces to the Schwarzschild-AdS and RN-AdS spacetime, if $b=Q=0$ and $b=0$, respectively.
These limiting cases verify that our charged Simpson–Visser-AdS regular metric interpolates smoothly between the well-known solutions.
\begin{figure}
\includegraphics[width=0.75\linewidth]{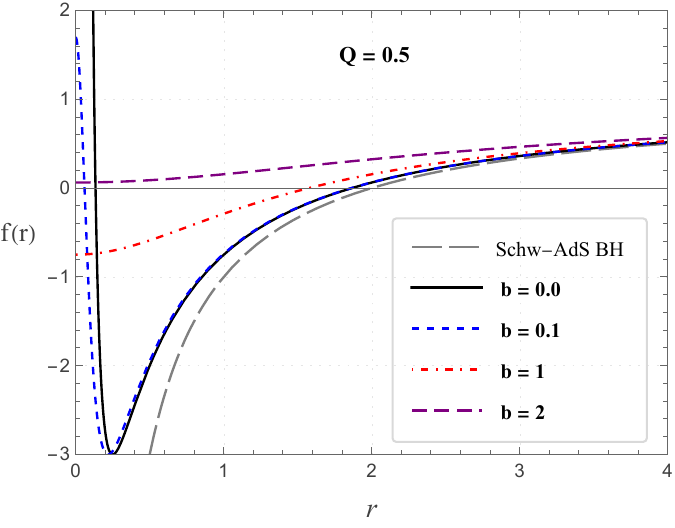}
\includegraphics[width=0.75\linewidth]{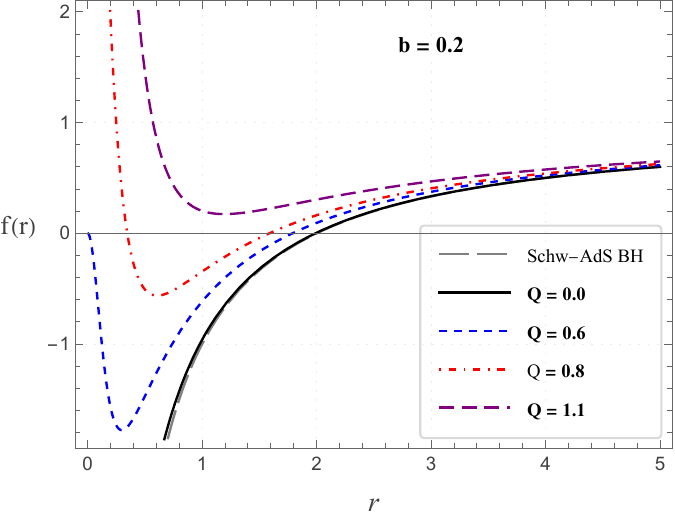}
\caption{Figures showing the radial behavior of metric function $f(r)$ along $r$, at $l=500$.}\label{Fig:function_f}
\end{figure}
%
\begin{figure*}[ht!]
\centering
\includegraphics[width=0.3\linewidth]{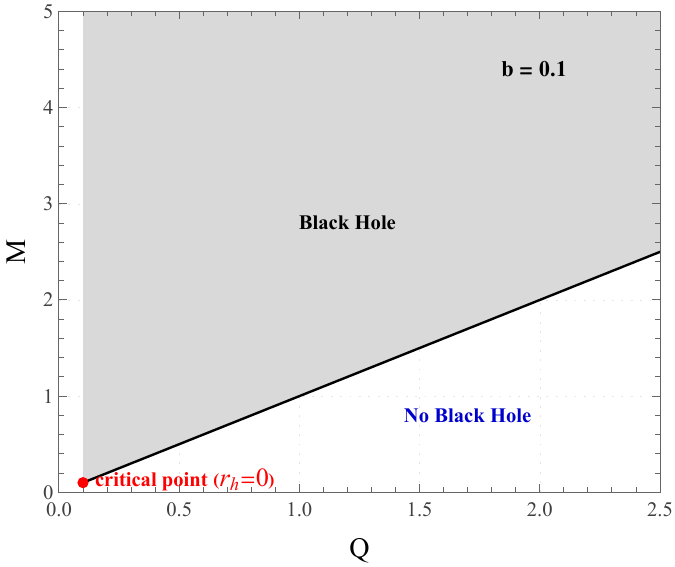}
\includegraphics[width=0.3\linewidth]{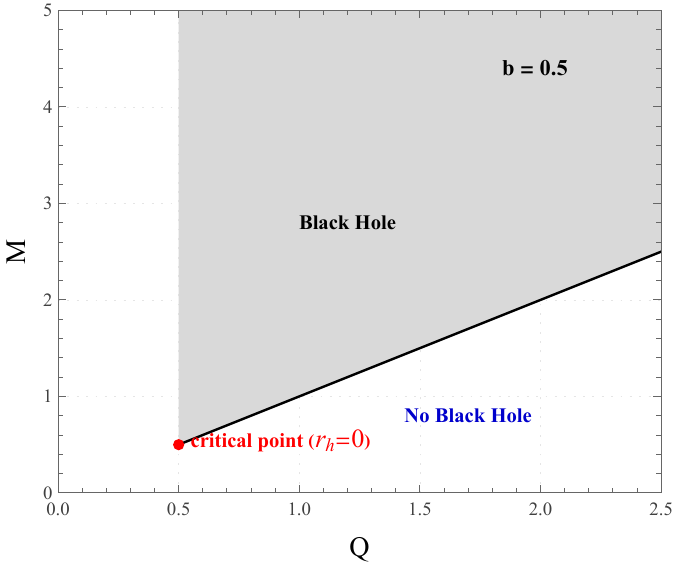}
\includegraphics[width=0.3\linewidth]{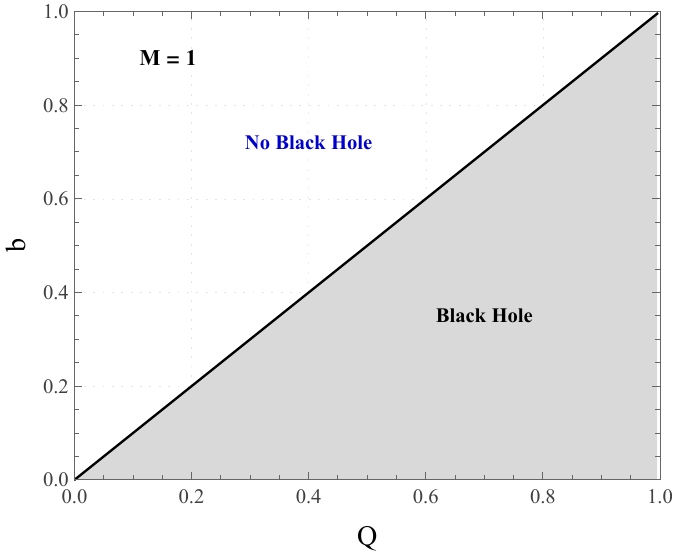}
\caption{Graphical illustration of the black hole phase diagram showing the region of black holes vs. no black hole in the $M$-$Q$ and $b$-$Q$ space at $l=500$.}\label{fig:BHregion}
\end{figure*}
\begin{figure*}[ht!]
\centering
\includegraphics[width=0.3\linewidth]{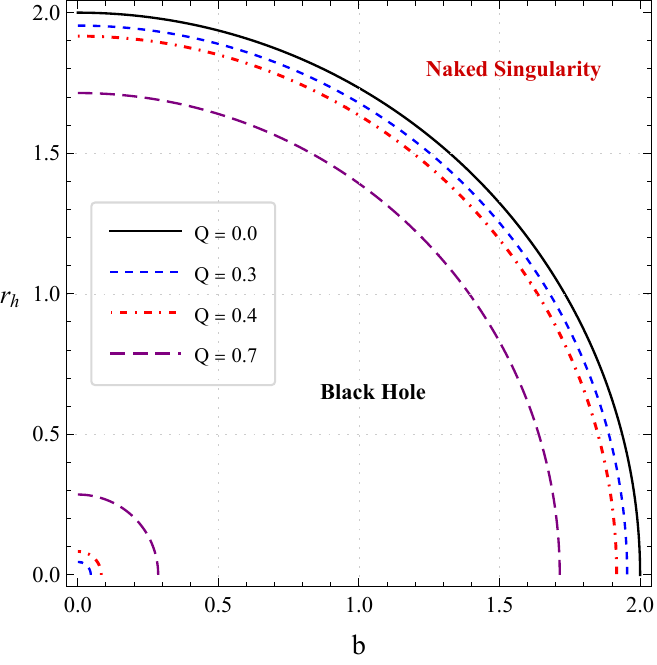}
\includegraphics[width=0.3\linewidth]{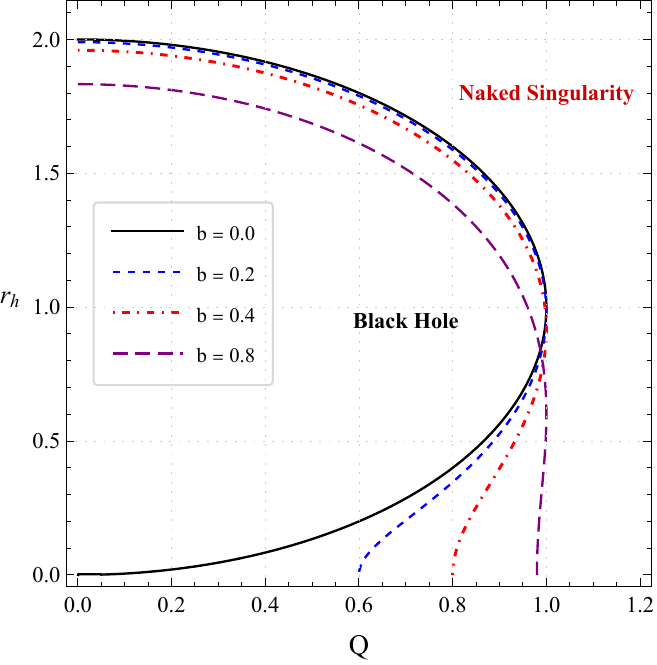}
\includegraphics[width=0.3\linewidth]{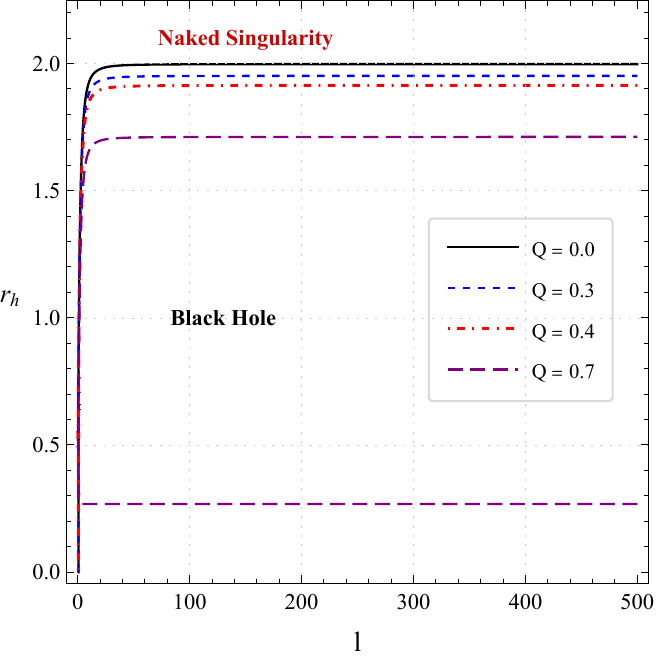}
\caption{Graphical interpretation of charged Simpson-Visser-AdS black hole horizons under variations in parametric values. In the left and right figures $l=500$, while in the middle $b= 0.1$.}\label{fig:horizons}
\end{figure*}

Figure~\ref{Fig:function_f} displays the radial profile of the metric function $f(r)$, which dictates the causal structure of the spacetime geometry. Compared with the RN-AdS black hole, the Schwarzschild-AdS black hole has lower values, and the fully generalized Simpson–Visser–AdS model shows systematically higher $f(r)$ values in the strong-field regime. The roots of $f(r) = 0$ map the spacetime's geometrical phase space, revealing transitions between three distinct topological classes: I) non-extremal configurations featuring distinct inner (Cauchy) and outer (event) horizons; II) extremal limits characterized by a single degenerate horizon; and III) strictly positive, horizonless geometries. 

Physically, both electromagnetic charge $Q$ and regularization parameter $b$ supply a repulsive contribution that counteracts the central gravitational mass. Consequently, increasing either parameter monotonically shrinks the outer event-horizon radius compared to the uncharged, singular black-hole limits. Critically, when these parameters exceed their respective extremal thresholds (e.g., at $Q = 1.1$ or $b = 2$), this effective repulsion dominates globally, providing $f(r) > 0$ for all $r$. Rather than exposing a classical naked singularity, this regime signals a phase transition in which the horizons vanish entirely, leaving a regular, horizonless compact object.

In Fig.~\ref{fig:BHregion}, we present the phase diagram of the parameter space, mapping the strict boundaries between geometries that admit an event horizon (the black hole phase) and those that do not (the horizonless regular regime). Compared to the standard RN-AdS black hole, the regularization parameter $b$ fundamentally restructures the regular black hole's geometry. Interestingly, $b$ shifts the critical point ($r_h = 0$) away from the black hole's origin. Physically, this shift is the defining topological signature of the regular geometry: it shows that the modified central core prevents horizon formation even at thresholds where standard general relativity would predict a classical black hole. Furthermore, since both $Q$ and $b$ supply effective repulsive contributions to the spacetime geometry, their coupled influence significantly compresses the allowable phase space for black hole existence, driving the system toward a regular, horizonless configuration at lower parametric thresholds.

To explore the horizons of the regular charged Simpson-Visser-AdS black hole, we make use of Eq.~\eqref{function_f}. Setting $f(r) = 0$ lets us plot the horizon radius numerically.\\
In Fig.~\ref{fig:horizons}, we mapped the parametric evolution of the inner and outer horizons as functions of the regularization parameter $b$ (left panel), the spacetime charge $Q$ (middle panel), and the AdS radius $l$ (right panel). The outer horizon $r_+$ initially expands rapidly with increasing $l$, a direct geometric consequence of weakening the negative cosmological constant and relaxing the background confinement. 
While, the electromagnetic charge $Q$ supplies a repulsive contribution to the effective potential. As $Q$ increases, $r_+$ squeezes while the inner horizon $r_-$ expands outward. This convergence restricts the physical horizon separation, suppresses the black hole surface area, and drives the spacetime toward an extremal state ($r_+ \to r_-$). 

The regularization parameter $b$ introduces a highly nonlinear geometric coupling to the electromagnetic charge. In the weak-charge regime, $b$ reinforces the effective central repulsion, working together with $Q$ to monotonically reduce the horizon radii. However, in the strong-charge regime, this scaling behavior strictly inverts. This topological reversal indicates that near extremality, the regularized core geometry strongly couples to the concentrated electromagnetic field, counteracting the standard RN horizon shrinkage and stabilizing the horizon structure against further contraction (see the middle panel of Fig.~\ref{fig:horizons} for details).
\section{Circular motion around charged Simpson-Visser-AdS black holes}\label{sec3: circular motion}
This section explores the characteristics of test-particle dynamics around static, spherically symmetric charged Simpson-Visser-AdS black holes of mass $m$, with the condition $p^{\mu}p_{\mu}=-m^2$.
Henceforth, the corresponding equation of motion for metric {given in Eq.~\eqref{metric}, can be obtained by using the Lagrangian density~\cite{2022MPLA...3750064K} as,
\begin{equation}\label{lagrangian1}
\mathcal{L}_p=\frac{1}{2}m g_{\mu \nu} {u}^{\mu} {u}^\nu.
\end{equation}
The corresponding time translation and rotational symmetry of the spacetime geometry related to conserved quantities can be determined through the Killing vectors
\begin{equation}
\xi_{(t)}^{\mu}\partial_{\mu}=\partial_{t} , \qquad
\xi_{(\phi)}^{\mu}\partial_{\mu}=\partial_{\phi},
\end{equation}
here $\xi_{(t)}^{\mu}=(1,~0,~0,~0)$ and $\xi_{(\phi)}^{\mu}=(0,~0,~0,~1)$, they are the specific energy $\mathcal{E}=E/m$ of the dynamic particles and its angular momentum $\mathcal{L}=L/m$.
In terms of the metric coefficients, they take the form
\begin{equation}\label{Energymom}
\mathcal{E}=-\frac{p_{t}}{m_0}=-f(r)\dot{t}, ~~~~
\mathcal{L}=\frac{ p_{\phi}}{m_0} = (r^2+b^2)\sin^2\theta \dot{\phi}.
\end{equation}
In the above expressions, the dot denotes the derivative with respect to proper time $\tau$. 
The expressions in Eq. \eqref{Energymom} can also be expressed as
\begin{equation}
\dot{t}=-\frac{\mathcal{E}}{f(r)},~~~~ \dot{\phi}=\frac{\mathcal{L}}{(r^2+b^2) \sin^2\theta}.
\end{equation}
Here, we consider the motion of a particle only in the equatorial plane ($\theta=\pi/2$, \text{so} $\dot{\theta}=0$)~\cite{2023EPJC...83..506K}. 
The normalization condition governing the equations of motion for a test particle takes the form
\begin{equation}\label{norm4vel}
g_{\mu \nu}u^{\mu}u^{\nu}=\epsilon \ .
\end{equation}
In the above expression, $\epsilon$ takes the values $-1$ and $0$ for massive and massless particles, respectively. The corresponding equation of motion for a particle with non-zero mass is governed by timelike geodesics. Henceforth, the equations of motion can easily be calculated utilizing the condition in Eq.~\eqref{norm4vel},
\begin{eqnarray}\label{eqmotionneutral}
\dot{r}^2&=&{\cal E}^2+f(r)\left(1+\frac{{\cal L}^2}{r^2+b^2}\right)\ ,
 \\
\dot{\theta}&=&\frac{1}{g_{\theta \theta}^2}\Big({\cal K}-\frac{{\cal L}^2}{\sin^2\theta}\Big)\ ,
\end{eqnarray}
here ${\cal K}$ represents the Carter constant. On making use of the normalization condition $g_{\mu\nu}\dot{x}^\mu \dot{x}^\nu = -1$, the equation of radial motion can be obtained as
\begin{eqnarray}
 \dot{r}^2={\cal E}^2-V_{\rm eff}\ ,
\end{eqnarray}
in which the effective potential takes the form
\begin{eqnarray}\label{effpotentail}
V_{\text{eff}} = f(r)\left(1+\frac{{\cal L}^2}{r^2+b^2}\right) .
\end{eqnarray}
\begin{figure*}[ht!]
\centering
  \includegraphics[width=0.325\linewidth]{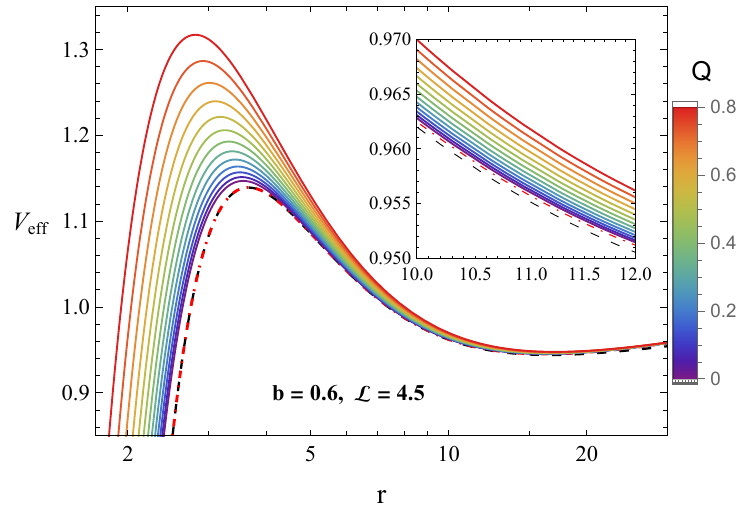}
    \includegraphics[width=0.325\linewidth]{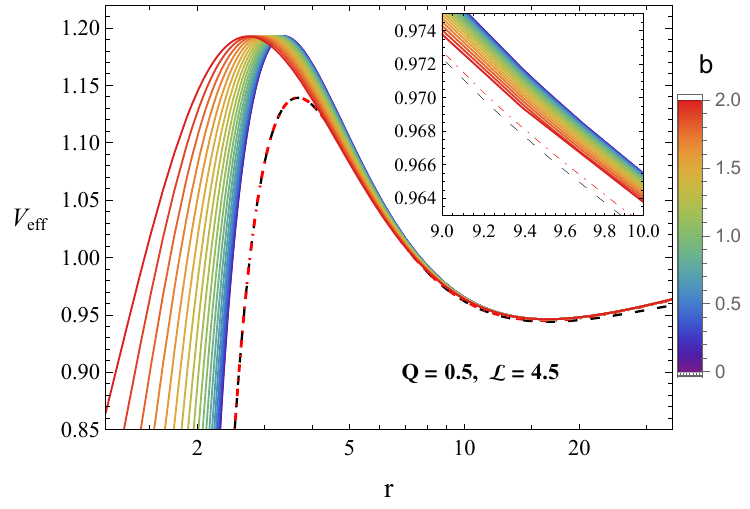}
      \includegraphics[width=0.325\linewidth]{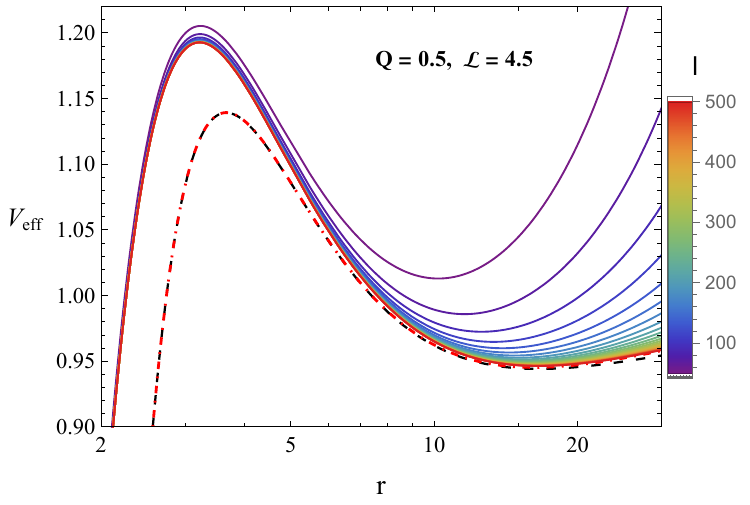}
   \caption{Graphical representation of a test particle's effective potential at $l=500$ (left and middle panels) and $b=1$ (right panel). The black dashed and red dotted-dashed curves correspond to the Schwarzschild and Schwarzschild-AdS black holes, respectively.}\label{plotVeff}
\end{figure*}

In Fig.~\ref{plotVeff}, we plot the radial dependence of the effective potential for various combinations of the black hole charge $Q$, regularization parameter $b$, and AdS radius $l$, compared with Schwarzschild (black dashed curves) and Schwarzschild-AdS (red dotted-dashed curves) black holes. The critical points $\partial_r V_{\text{eff}} = 0$ map the orbital phase space, where local minima and maxima strictly govern the domains of stable and unstable circular orbits, respectively.

The Schwarzschild geometry provides a standard potential well for stable orbits; the Simpson–Visser–AdS spacetime fundamentally restructures the strong-field regime. Specifically, the electromagnetic charge $Q$ provides an effective repulsion that raises the centrifugal barrier; this amplifies the instability maxima associated with unstable circular orbits and shifts the stable orbital domains outward. The AdS radius $l$ exerts a competing influence, because the negative cosmological constant ($\Lambda = -3/l^2$) acts as an inward harmonic confinement, and increasing $l$ monotonically dilutes the background curvature. 
As $l$ expands toward the asymptotically flat limit (e.g., $l = 500$), the cosmological confinement decays, suppressing the inner instability barrier and deepening the primary potential well. In this extended regime, the regularized potential converges remarkably to the classical Schwarzschild stability minimum, showing that a weakened AdS background effectively neutralizes particle instability near the regularized throat while preserving long-range bound trajectories. Moreover, $b$ shifts the circular orbits closer to the black hole.
\subsection{Circular orbits}\label{subsec: circular orbits}
To determine circular geodesics, one can use the simultaneous conditions:
\beq
V_{\rm eff}(r) = \mathcal{E}^2, \quad \frac{\partial V_{\rm eff}(r)}{\partial r} = 0.
\eeq
From these conditions, the specific angular momentum $\mathcal{L}$ and energy $\mathcal{E}$ for circular orbits can be derived as:
\begin{equation}
  \mathcal{E}^{2}=\frac{2r f(r)^{2}}{\mathcal{D}},\qquad
  \mathcal{L}^{2}=\frac{(r^2+b^2)^2f^{\prime}(r)}{\mathcal{D}},
  \label{eq:EL}
\end{equation}
with 
\beq 
\mathcal{D}\equiv 2rf(r)-(r^2+b^2)f^{\prime}(r),
\eeq
In the above expressions, $f^{\prime}(r)$ denotes the derivative of the metric function $f(r)$ with respect to $r$.
%
\subsection{Angular momentum and energy}\label{subsec: angular momentum}
In this subsection, we examine the angular momentum and energy of a particle orbiting a regular charged Simpson-Visser-AdS black hole.

\begin{figure*}[ht!]
    \includegraphics[width=0.3\textwidth]{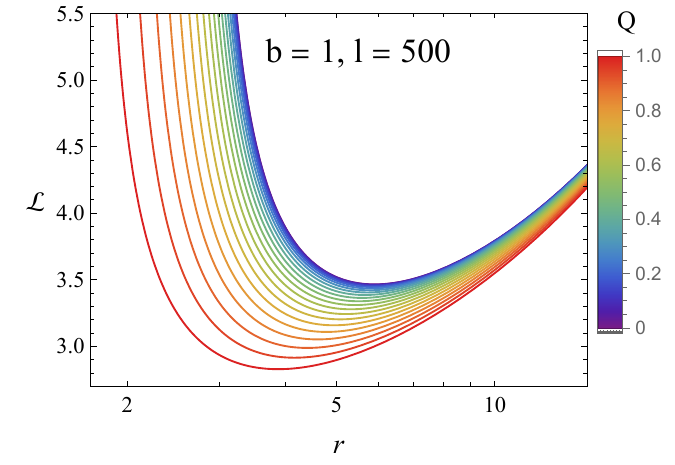}
    \includegraphics[width=0.3\textwidth]{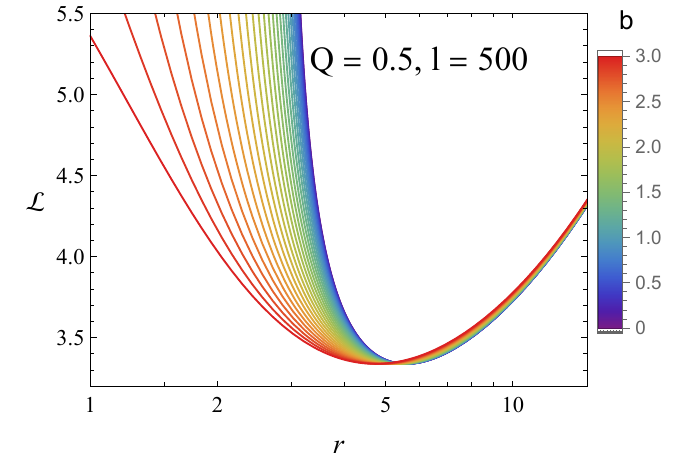}
    \includegraphics[width=0.3\textwidth]{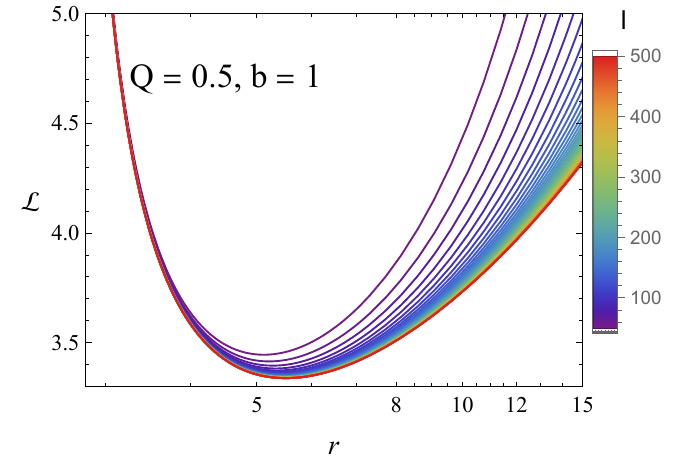}
    \includegraphics[width=0.3\textwidth]{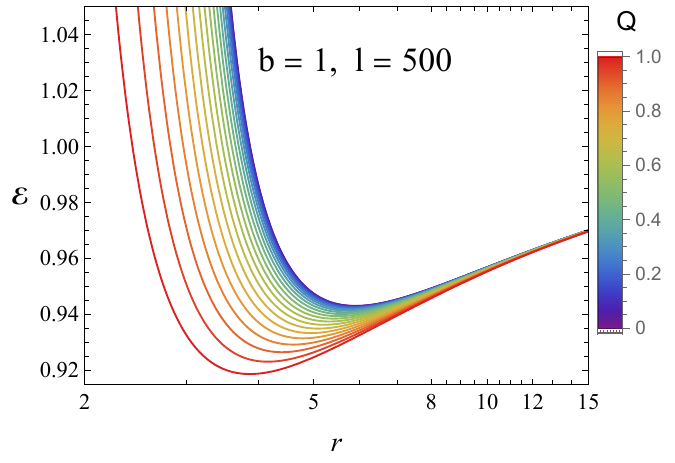}
    \includegraphics[width=0.3\textwidth]{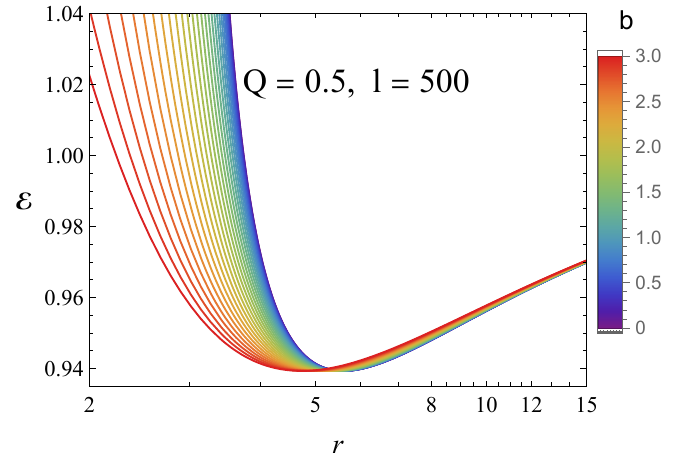}
    \includegraphics[width=0.3\textwidth]{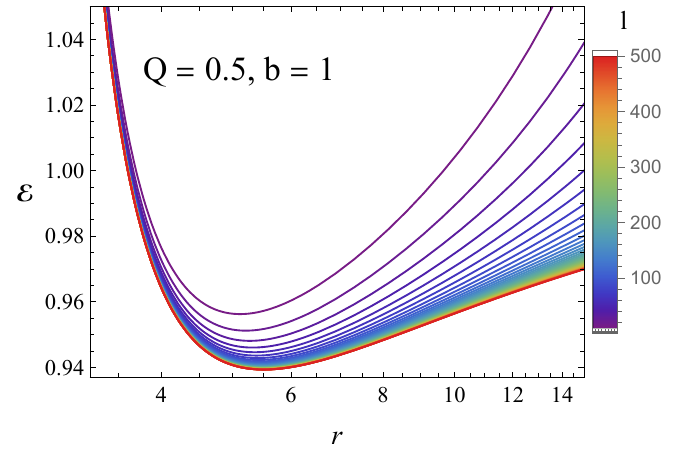}
\caption{Radial profiles of a particle's specific angular momentum (top row) and energy (bottom row) for various discrete values of the black hole parameters.}\label{Fig:A.mom_eng}
\end{figure*}
Figure~\ref{Fig:A.mom_eng} shows the radial profiles of the specific angular momentum ($\mathcal{L}$) and specific energy ($\mathcal{E}$) for test particles in circular orbits. The global minimum of each curve distinctly marks the location and kinematic threshold of the ISCO. We observe that the black hole charge $Q$ and the AdS radius $l$ monotonically shift the ISCO inward to smaller radii while simultaneously lowering the curves. 

The regularization parameter $b$ has a similar effect on the kinematic thresholds, reducing the required $\mathcal{E}$ and $\mathcal{L}$ for orbital stability, though it induces only marginal shifts in the ISCO radius. 
Physically, this behavior indicates that the regularized core's modified spacetime geometry softens the effective gravitational potential in the strong-field regime. As the effective gravitational pull weakens, the kinematic barrier is reduced, allowing test particles to sustain stable circular orbits closer to the black hole with lower specific energy and angular momentum. Furthermore, because the negative cosmological constant of the AdS background acts as a confining harmonic potential, weakening this background at higher values of the AdS radius (e.g., $l=500$) yields the absolute lowest minima for both $\mathcal{E}$ and $\mathcal{L}$, recovering behavior remarkably closer to the asymptotically flat regime.
\subsection{Innermost stable circular orbits}\label{subsec: ISCO}
Stable circular orbits occur at the radius $r = r_{min}$, where the particles' minimum energy and angular momentum correspond to circular orbits. For the innermost stable circular orbits (ISCOs), the conditions $\partial_r V_{\rm eff}=0,$ and $\partial_{rr}V_{\rm eff}=0$ must be satisfied. After some algebraic simplifications, the expression of the ISCO radius took the form 
\begin{eqnarray}\nonumber
r_{\rm ISCO}&=&\frac{2 r f(r)}{\left(b^2+r^2\right)^3 \left(\left(b^2+r^2\right) f^{\prime}(r)-2 r f(r)\right)} \\ \nonumber
        &&\times \Big[2 \left(b^2-3 r^2\right) f(r)^2 +4 r \left(b^2+r^2\right) f(r) f^{\prime}(r) \\\label{ISCO}
        &&-\left(b^2+r^2\right)^3 f^{\prime \prime}(r)\Big].
\end{eqnarray}
%
\begin{figure*}[ht!]
\centering
    \includegraphics[width=0.28\linewidth]{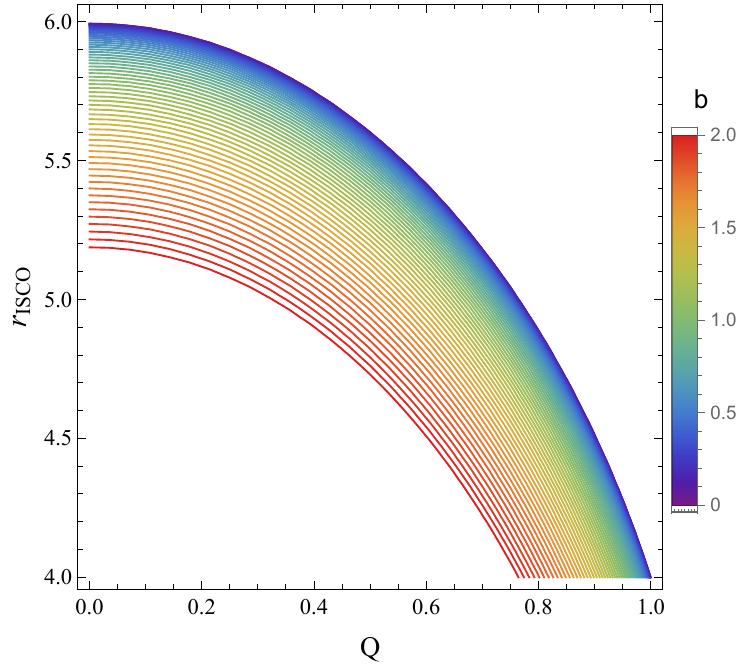}
    \includegraphics[width=0.28\linewidth]{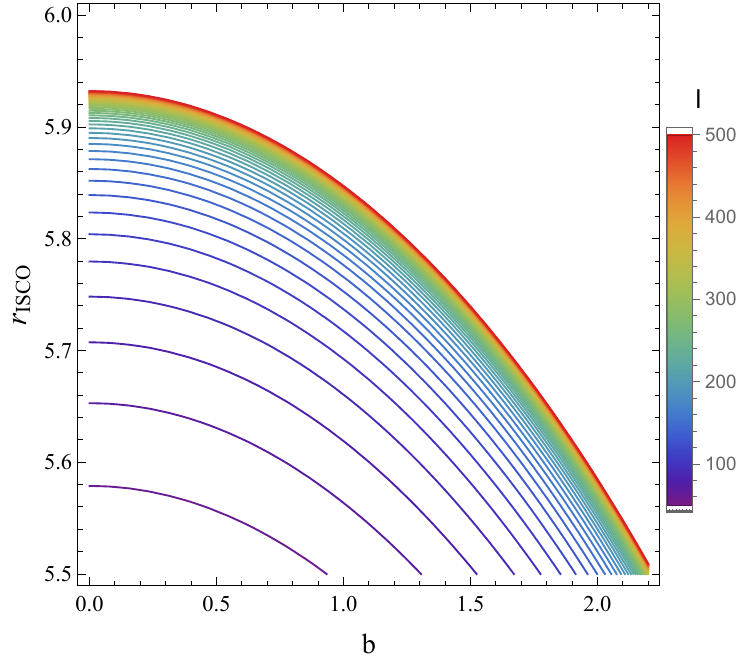}
    \includegraphics[width=0.28\linewidth]{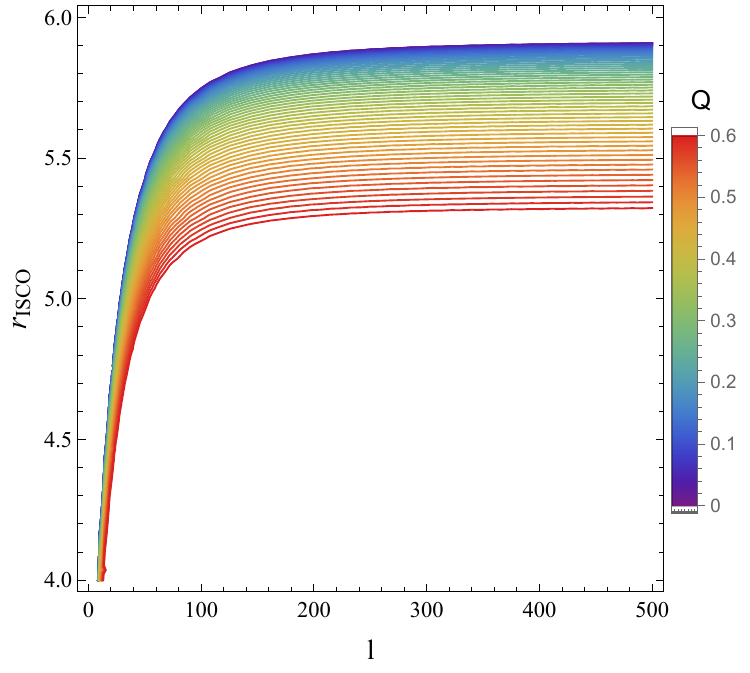}
\caption{Graphical illustration of the ISCO radius vs. black hole charge $Q$, regularization parameter $b$, and AdS radius $l$ at $l=500$ (left), $Q=0.2$ (middle), and $b=1$ (right).\label{fig:ISCO}}
  \end{figure*}

Figure~\ref{fig:ISCO} illustrates the parametric dependence of the ISCO radius $r_{\rm ISCO}$ on the regularization parameter $b$, the AdS radius $l$, and the spacetime charge $Q$. Our results show that $r_{\rm ISCO}$ decreases monotonically as both $Q$ and $b$ increase. Dynamically, this inward shift occurs because the electromagnetic charge supplies a repulsive contribution to the effective potential, while the regularization parameter modifies the central geometry, effectively softening the strong-field gravitational gradient near the core. 

Moreover, the ISCO response to the AdS radius $l$ exhibits a distinct, highly nonlinear scaling. In the deep AdS regime (small $l$), the strong confining effect of the negative cosmological constant couples heavily to the orbital dynamics, driving pronounced shifts in $r_{\rm ISCO}$. However, as $l$ increases, the spacetime smoothly dilutes toward asymptotic flatness ($\Lambda \to 0$). In this regime, cosmological confinement decays rapidly, causing the ISCO radius to decouple from $l$ and converge asymptotically to the flat-space limit characteristic of a charged Simpson––Visser black hole.
\section{Fundamental frequencies and QPOs}\label{sec4:Fund_frequencies and QPOs}
This section explores the fundamental frequencies governing test-particle motion near a charged Simpson--Visser--AdS black hole. To evaluate small perturbations near stable circular orbits, we investigate the orbital (Keplerian) frequency, along with the radial and vertical epicyclic frequencies. Dynamical frequencies are vital for understanding matter behavior in the intense gravitational environment around black holes and are integral to interpreting observable signals, including twin-peak QPOs. Investigating their characteristics allows us to infer physically significant constraints on the possible values of the upper and lower QPO frequencies observed in astrophysical systems.
\subsection{Harmonic oscillations}\label{subsec: Harmonic_Oscs}
In this subsection, we examine the fundamental frequencies arising from the oscillatory motion of test particles orbiting a charged Simpson––Visser––AdS black hole.
Test particles on stable circular orbits in the equatorial plane undergo small oscillations along the radial, angular, and vertical directions. We obtain the fundamental radial and vertical frequencies by perturbing the circular orbit along the radial ($r\to r_0+\delta r$) and vertical ($\theta\to \theta_0+\delta\theta$) directions, respectively.

Expanding the effective potential in terms of $r$ and $\theta$, we obtain
\begin{eqnarray}\label{Vexpand}
\nonumber
V_{\rm eff}(r, \theta) &=& V_{\rm eff}(r_0,\theta_0) +\delta r\,\partial_r V_{\rm eff}(r,\theta)\Big|_{x_0} 
\\\nonumber
&& +\delta\theta\, \partial_\theta V_{\rm eff}(r,\theta)\Big|_{x_0} +\frac{1}{2}\delta r^2\,\partial_r^2 V_{\rm eff}(r,\theta)\Big|_{x_0}
\\\nonumber
&& +\frac{1}{2}\delta\theta^2\,\partial_\theta^2 V_{\rm eff}(r,\theta)\Big|_{x_0} +\delta r\,\delta\theta\,\partial_r\partial_\theta V_{\rm eff}(r,\theta)\Big|_{x_0}\\&&
+{\cal O}\left(\delta r^3,\delta\theta^3\right)\, ,
\end{eqnarray}
where $x_0= (r_0, \theta_{0})$. It may be seen that the first term of equation~\eqref{Vexpand} becomes null given that $V_{\rm eff}(r_0,\theta_0)=0$. Furthermore, the subsequent and third terms vanish because the effective potential is stable ($ \partial_r V_{\rm eff}=\partial_\theta V_{\rm eff}= 0$). The only terms involving second derivatives of the effective potential depend on $r$ and $\theta$. To obtain the physical values recorded by a remote observer in the equation of motion, we substitute the relation connecting the affine parameter with the time derivative ($dt/d\lambda=u^t$). Using basic algebraic computations and considering the preceding details, we can straightforwardly derive the harmonic oscillatory equations for displacements $\delta r$ and $\delta\theta$, as
\bea
&& \Omega_\phi = \sqrt{\frac{f'(r)}{2r}}, \\
&&\Omega_r = \sqrt{\frac{f(r)}{2E^2} \frac{\partial^2 V_{\text{eff}}}{\partial r^2}}\bigg|_{\theta=\pi/2},\\
&&\Omega_\theta = \sqrt{\frac{f(r)}{2E^2} \frac{\partial^2 V_{\text{eff}}}{\partial \theta^2}}\bigg|_{\theta=\pi/2}.
\eea
The straightforward differentiation of the effective potential results in the mathematical equations for the radial frequency profile as follows:
\bea \nonumber
\Omega_r^2 &= &\frac{f^2(r)}{2 \EE^2} \bigg[ f^{\prime\prime}(r)\left(1+\frac{ \LL^2}{r^2+b^2}\right)  \\ 
            && - \frac{4r \LL^2 f^{\prime}(r)}{(r^2+b^2)^2}- \frac{2 \LL^2 f(r)(b^2-3r^2)}{(r^2+b^2)^3} \bigg]. \label{eq:omega1}
\eea
On eliminating $\mathcal{E}^{2}$ and $\mathcal{L}^{2}$ through Eq.~\eqref{eq:EL}, Eq.~\eqref{eq:omega1} yields the closed form
\begin{equation}
  \Omega_{r}^{2}=\tfrac{1}{2}f(r) f^{\prime\prime}(r) -(f^{\prime}(r))^2
  +\frac{\bigl(3r^{2}-b^{2}\bigr)}{2r\bigl(r^{2}+b^{2}\bigr)}f(r) f^{\prime}(r).  \label{eq:omega2}
\end{equation}

On substituting $b=Q=0$, and $l\to\infty$, Eq.~\eqref{eq:omega2} reduces to $\Omega_{r}^{2}=r^{-3}(1-6/r)$ for which we verified numerically agrees with the
Schwarzschild result to within $2\times10^{-16}$ at $r=6.5$--$50\,M$, before any fit was performed. Since the metric in Eq.~\eqref{metric} with \eqref{function_f} is static, so $\Omega_{\theta}=\Omega_{\varphi}$ are identical. Therefore,  the corresponding geometry cannot produce nodal precession; we hence confront it only with the two high-frequency peaks.\\
\begin{figure}
\centering 
\includegraphics[width=0.7\linewidth]{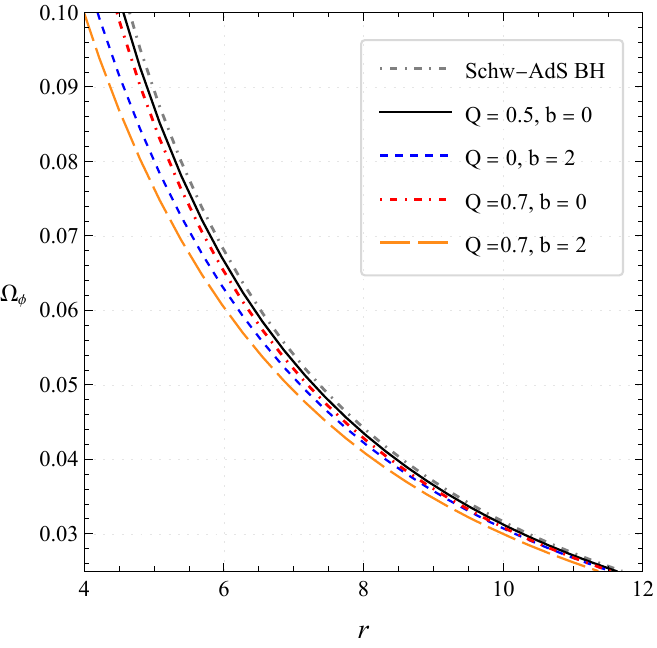}
\includegraphics[width=0.7\linewidth]{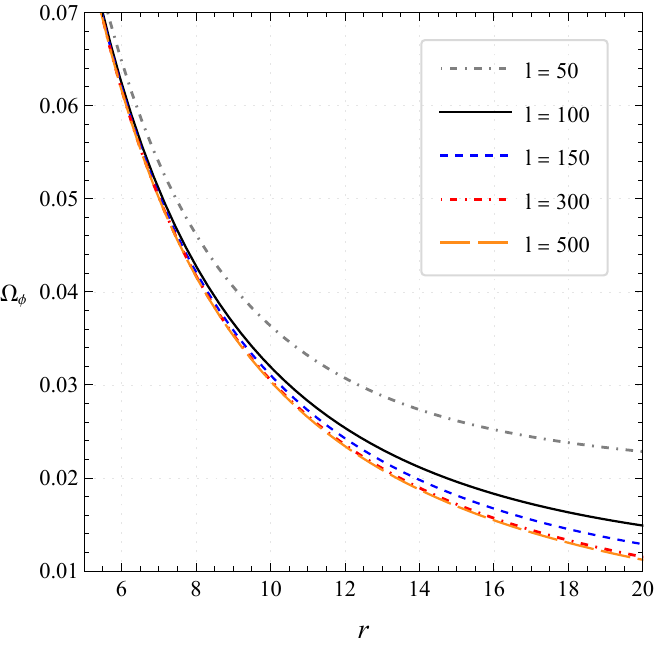}
\caption{Radial dependence of Keplerian frequencies of test particles around a charged Simpson-Visser black hole. The top panel is plotted at $l=500$ while the bottom panel is plotted at $Q=0.5$ and $b=2$.} \label{Fig:Kpfrequencies}
\end{figure}
\begin{figure}
\centering 
\includegraphics[width=0.7\linewidth]{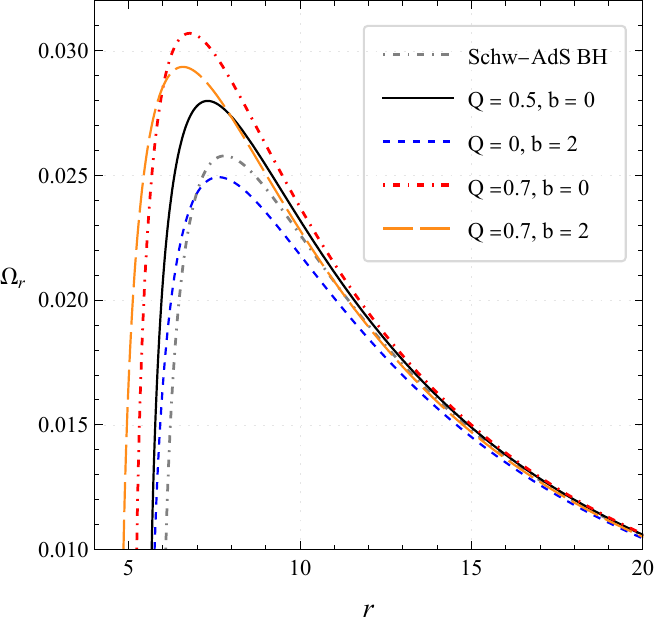}
\includegraphics[width=0.7\linewidth]{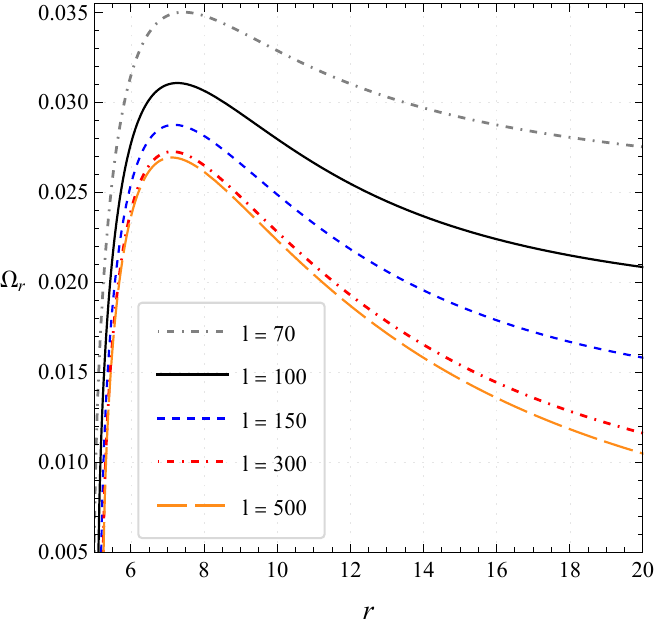}
\caption{Radial dependence of radial frequencies of test particles around a charged Simpson-Visser black hole. The top panel is plotted at $l=500$ while the bottom panel is plotted at $Q=0.5$ and $b=2$.} \label{Fig:Rdfrequencies}
\end{figure}
Figures~\ref{Fig:Kpfrequencies} and~\ref{Fig:Rdfrequencies} illustrate the radial profiles of the Keplerian ($\Omega_\phi$) and radial epicyclic ($\Omega_r$) frequencies for varying values of the spacetime charge $Q$, regularization parameter $b$, and AdS radius $l$. As expected from standard orbital kinematics, the Keplerian frequency decays monotonically with radial distance, which supports previous findings. \\
However, in the strong-field regime, both the electromagnetic charge $Q$ and the regularization parameter $b$ introduce effective repulsive corrections to the central gravitational well. This repulsion lowers the required orbital velocity, systematically suppressing $\Omega_\phi$ in the charged Simpson–Visser–AdS spacetime relative to the singular Schwarzschild-AdS black hole. While the modifications from $Q$ and $b$ dominate the inner geometry, the AdS radius $l$ becomes increasingly influential at larger radii, governing the deviation from asymptotic flatness and further suppressing the Keplerian frequency in the outer domains.

Similarly, in the radial epicyclic frequency $\Omega_r$, the regularization parameter $b$ and AdS radius $l$ reduce the radial frequency profile as $r$ increases. Unlike the Keplerian frequency, the black hole charge $Q$ increases the radial frequency. The curves exhibit a well-defined maximum, whose position marks the transition region between the strongly curved inner geometry and the weak-field regime.
We found that $Q$ uniformly shifts the entire profile upward, increases the peak value of $\Omega_r$, and slightly shifts the maximum inward toward the black hole. Moreover, $b$ and $l$ cause a uniform downward shift in the maxima. Furthermore, the Simpson-Visser-AdS black hole has a smaller $\Omega_r$ profile than Schwarzschild-AdS and charged Simpson-Visser-AdS black holes.

For theoretical considerations, we can formulate the basic frequencies in Hertz. In the International System of Units, the speed of light in a vacuum is $c=3\times 10^8 \rm m/sec$, and the gravitational constant is $G=6.67 \times 10^{-11}\rm m^3/(kg^2\cdot sec)$. Consequently, the corresponding frequencies take the following form
\begin{equation}\label{nu_i}
\nu_{i} = \frac{1}{2\pi}\frac{c^3}{GM} \Omega_{i} , [{\rm Hz}]\ .
\end{equation}
In the aforementioned equation~\eqref{nu_i}, $i \in \{r, \theta, \phi\}$.
In the Relativistic Precession (RP) model, the upper and lower HF-QPO frequencies ($\nu_U, \nu_L$) are explicitly given by $\nu_U = \nu_\phi$ and $\nu_L = \nu_\phi - \nu_r$.
\begin{figure*}
\centering
\includegraphics[width=0.31\linewidth]{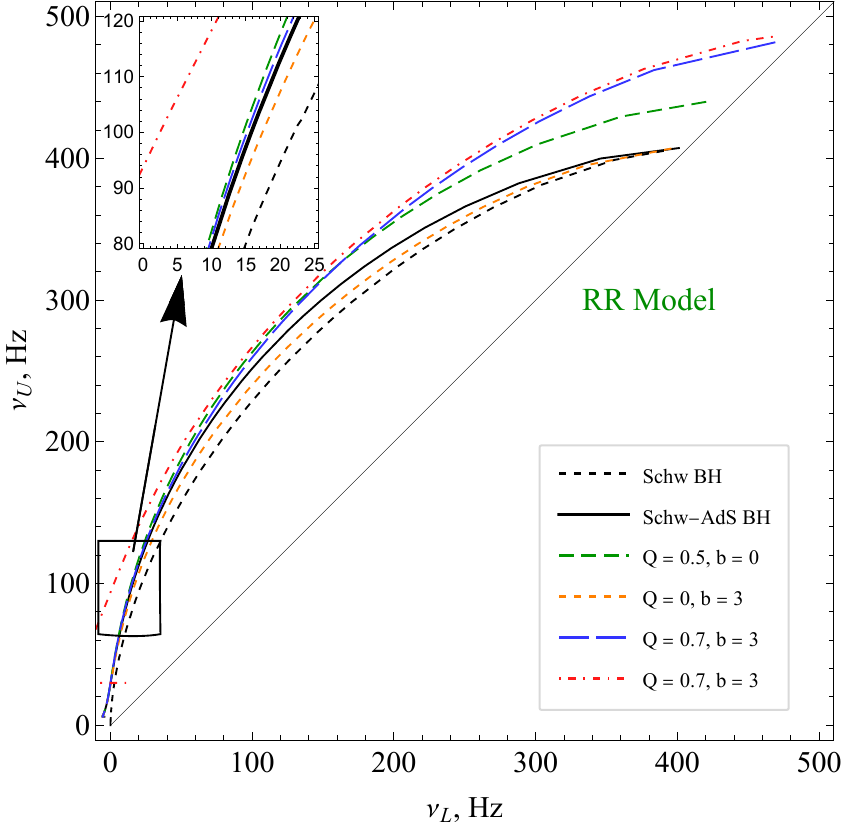}
\includegraphics[width=0.31\linewidth]{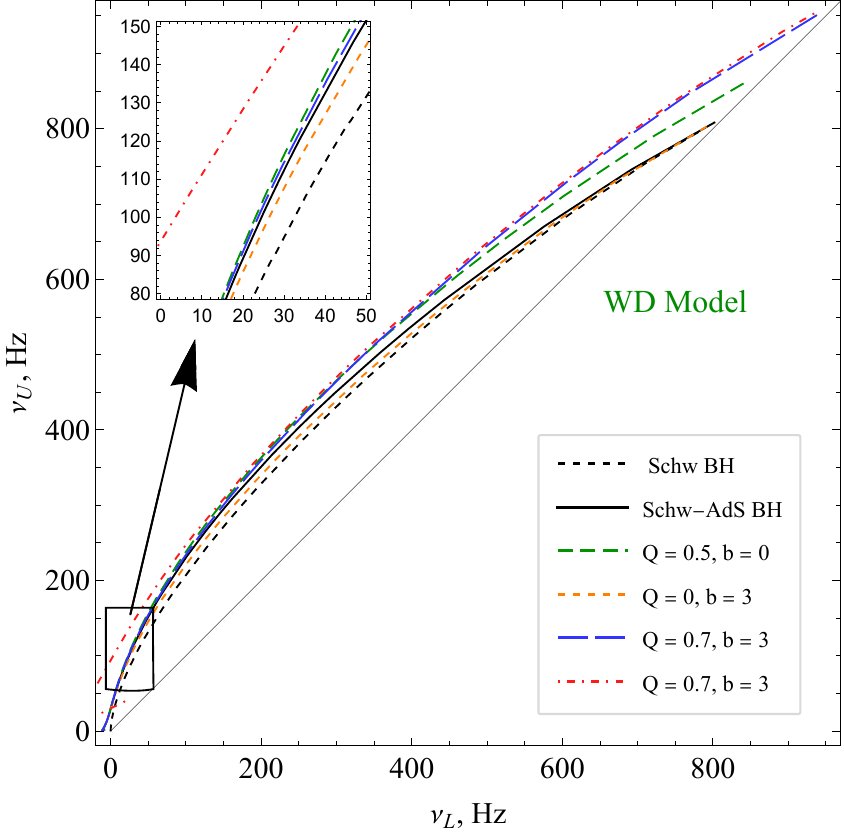}
\vspace{0.3cm}
\includegraphics[width=0.31\linewidth]{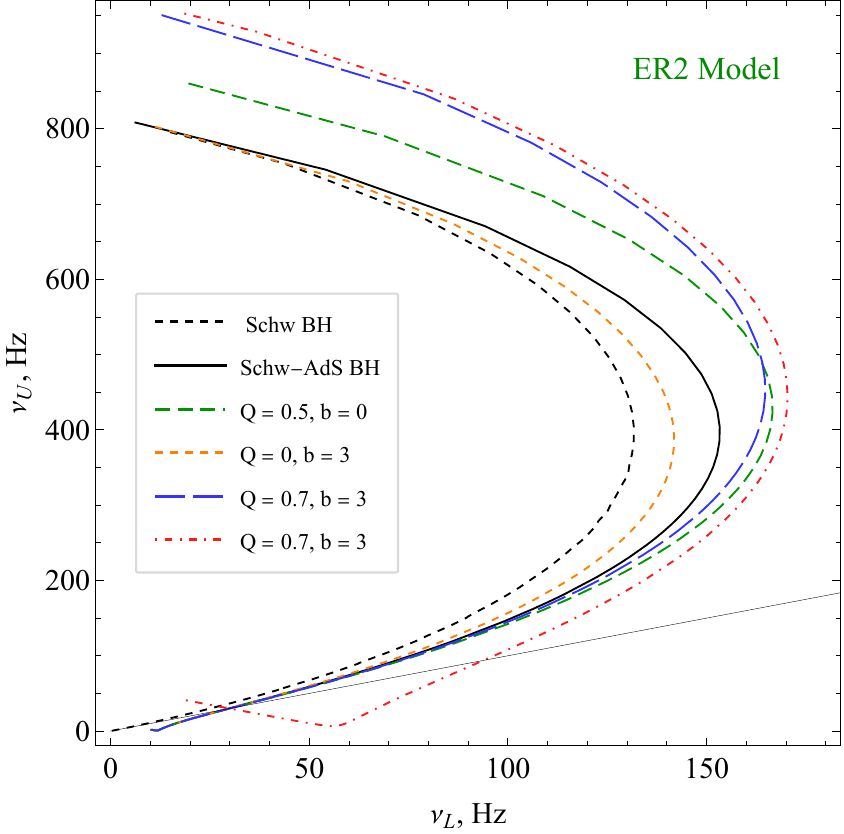}
\includegraphics[width=0.31\linewidth]{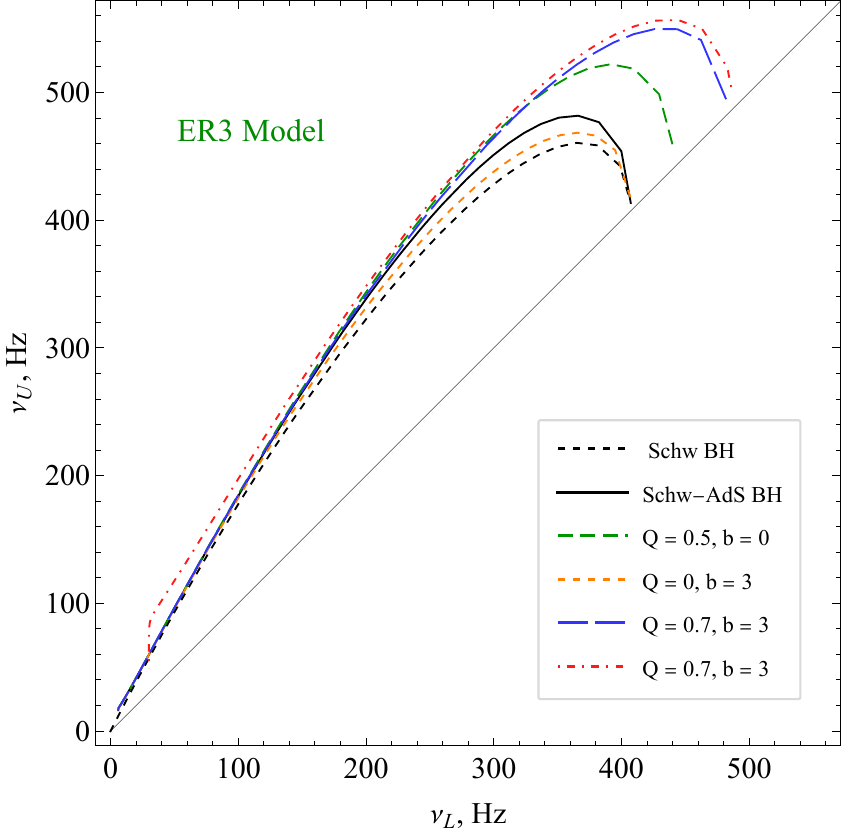}
\includegraphics[width=0.31\linewidth]{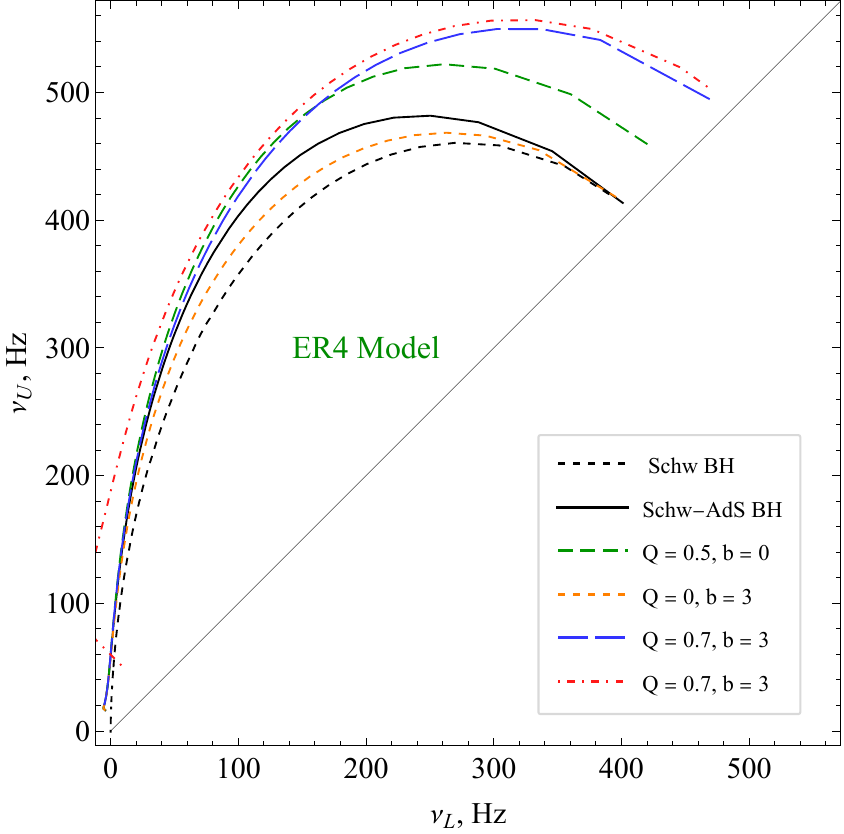}
\caption{Relations between the upper and lower frequencies of twin peak QPOs in the RP, WD, and ER2-4 models around charged Simpson-Visser-AdS black holes at $l=1000$, except for the red dotted-dashed curves, which are plotted at $l=200$.}\label{QPOs}
\end{figure*}
\subsection{Quasi-periodic oscillations}\label{subsec:QPOs}
Quasi-periodic oscillations (QPOs) are among the most compelling astrophysical phenomena observed near compact objects, including black holes, neutron stars, white dwarfs, and their binary companions~\cite{Ingram2010MNRAS,Zdunik2000AA}. While QPOs occur across various compact systems, observational data show they are most frequently detected in neutron star binaries, with a smaller subset linked to black holes and white dwarfs. The leading theoretical framework attributes these signals to (quasi)harmonic oscillations of accreting matter within strong-field regimes. Consequently, QPO frequencies serve as high-precision probes of spacetime geometry, enabling empirical constraints on central object properties. Observations of QPO frequencies in black hole systems may constrain the parameters of Simpson-Visser models, facilitating discrimination among different black hole scenarios.

In the context of MOG theories ~\cite{2024EPJC...84..964J,2025EPJC...85.1029N} and exotic field models, QPOs provide critical diagnostic value: observed frequency ratios and peak profiles directly constrain both local matter compositions~\cite{2025PDU....5002110S,2024EPJC...84.1114R} and underlying gravitational dynamics. Analyzing these oscillations therefore bridges accretion physics with observational relativity, offering a concrete mechanism to test phantom fields, evaluate alternative gravity frameworks, and differentiate between distinct black hole metrics.

In this work, we investigate twin-peak QPO frequencies in the spacetime of a charged Simpson–Visser–AdS black hole and compare our results directly with the Schwarzschild–AdS metric. Following standard theoretical frameworks~\cite{2021EPJC...81.1067S}, we analyze three principal QPO models: (I) Relativistic Precession (RP) Model~\cite{Stella1999ApJ}: Assumes the upper and lower frequencies correspond to the azimuthal and periastron precession frequencies, defined as $\nu_U = \nu_\phi$ and $\nu_L = \nu_\phi - \nu_r$.
(II) Warped Disc (WD) Model: Attributes QPOs to oscillations of test particles within a thin accretion disc, where $\nu_U = 2\nu_\phi - \nu_r$ and $\nu_L = 2(\nu_\phi - \nu_r)$.
(III) Epicyclic Resonance (ER) Models: Assume a geometrically thick accretion disc where QPOs stem from resonant geodetic oscillations of radiating particles. We evaluate three specific variants, ER2: $\nu_U = 2\nu_\theta - \nu_r$ and $\nu_L = \nu_r$; ER3: $\nu_U = \nu_\theta + \nu_r$ and $\nu_L = \nu_\theta$; and ER4: $\nu_U = \nu_\theta + \nu_r$ and $\nu_L = \nu_\theta - \nu_r$.

Figure~\ref{QPOs} shows the trajectories of the upper ($\nu_U$) and lower ($\nu_L$) epicyclic frequencies in the $\nu_U$--$\nu_L$ plane for a charged Simpson–Visser–AdS black hole, compared with standard Schwarzschild and Schwarzschild-AdS spacetimes. We consider a stellar-mass black hole with $M \approx 5.4 \pm 0.3 M_\odot$ (GRO~J1655$-$40). The curves trace the radial domain extending from the ISCO outward across the RP, WD, and ER2–4 models for various combinations of the charge $Q$, regularization parameter $b$, and AdS radius $l$. Lines of small-integer frequency ratios ($3:2$, $4:3$, $5:4$) mark potential orbital resonances, alongside the diagonal degeneracy limit ($\nu_U: \nu_L = 1:1$). Below this boundary, the twin-peak QPOs cannot physically form; emission radii approaching this $1:1$ line exhibit degenerate epicyclic frequencies that merge into a single observed peak.

In the low-frequency regime, the regularization parameter $b$ systematically suppresses the frequency tracks relative to the Schwarzschild-AdS black hole across all applied models. Physically, this suppression originates from the modified geometry near the regularized core, which softens the effective gravitational potential and weakens the epicyclic restoring forces. Moreover, the AdS radius $l$ considerably modulates the QPO spectra: smaller values of $l$ intensify the negative cosmological background confinement ($\Lambda = -3/l^2$), strengthening the epicyclic restoring forces to generate higher peak QPO frequencies. Conversely, expanding $l$ relaxes this spatial confinement toward asymptotic flatness, softening orbital epicyclic oscillations and reducing the peak QPO frequencies across all models. 

In contrast, the electromagnetic charge $Q$ provides a competing repulsive contribution that stiffens the orbital dynamics. This elevates the low-frequency tracks and consistently drives the peak frequencies to systematically higher values than those permitted in the singular, uncharged limits. This strict dynamical tension between $b$ and $Q$ offers a compelling phenomenological signature for observational discrimination. As regular spacetime corrections soften the QPO spectrum while electromagnetic charge stiffens it, these opposing mechanisms provide a rigorous theoretical framework for constraining non-singular black-hole parameters using precision X-ray timing data from microquasars.

Our analysis shows that the regularization parameter $b$ and AdS radius $l$ produce different twin-peak QPOs, as $b$ produces higher QPO frequencies than in a Schwarzschild black hole. Our findings show a non-linear resonance shift when $Q$ and $l$ are present simultaneously. For a given $b$, the $3:2$ resonance ratio occurs at a different radial coordinate $r$, unlike the uncharged Simpson-Visser case.

Additionally, unlike previous investigations of black holes in GR, the electromagnetic charge, Simpson-Visser coupling, and AdS radius $l$ produce an updated ISCO. This suggests a potential observational signature: a specific range of QPO frequencies that can only occur when the vector field $\phi_\mu$, $b$, $Q$, and $l$ are all non-zero. This may allow future X-ray timing missions to distinguish between geometric and field-theoretical deviations from Einstein gravity.
\subsection{Statistical constraints from HF-QPOs}\label{sec:qpo}
To assess the compatibility of the charged Simpson––Visser––AdS spacetime with the twin-peak HF-QPOs observed in black-hole X-ray binaries, we examine two microquasar benchmarks, \textbf{GRO~J1655$-$40} and \textbf{XTE~J1550$-$564}, which exhibit pronounced HF-QPOs interpreted via the relativistic-precession (RP) model. RP is the microquasar HF-QPO fitting model most widely used because of its strong observational constraints.

In this segment, we work in the equatorial plane, use geometrized units with $G=c=1$, and express all lengths in terms of the mass parameter $M$, making $r$, $b$, $Q$, and $l$ dimensionless.
\subsubsection{Data and statistical procedure}
We consider the two black hole binary systems for which a twin-peak HF-QPO detection is accompanied by a dynamical mass measurement: GRO~J1655$-$40 ($\nu_{U}=450\pm3$~Hz, $\nu_{L}=300\pm5$~Hz, $M=5.4\pm0.3\,M_{\odot}$) and XTE~J1550$-$564 ($\nu_{U}=276\pm3$~Hz, $\nu_{L}=184\pm5$~Hz, $M=9.1\pm0.6\,M_{\odot}$)~\cite{Strohmayer_2001ApJ...552L..49S,Beer2002MNRAS.331..351B,Orosz2011ApJ...742...84O}. To acquire estimations on parameters such as the peak frequencies of QPOs observed in the microquasars, we perform the $\chi^2$ analysis \cite{2015EPJC...75..162B}:
\begin{equation}
\chi^{2} = \sum_{i \in \{U, L\}} \frac{(\nu_{i}^{mod}-\nu_{i}^{obs})^{2}%
}{\sigma_{i}^2}  +\left(\frac{M-M^{\rm obs}}{\sigma_{M}}\right)^{2}.
\label{eq:chi2}
\end{equation}
The last term implements the dynamical mass as a Gaussian prior rather than as a fixed input. Consequently, every source provides three data points. The emission radius $r_{i}$ and the mass $M_{i}$ are considered as free parameters, allowing for a fit at constant $(Q,b,l)$ to provide
$\nu_{\rm dof}=3-2=1$.

We perform the minimization via a bounded differential-evolution global search followed by local gradient refinement, using a fixed random seed to ensure reproducibility. Configurations are permitted only if $f>0$, $\mathcal{D}>0$, $\mathcal{E}^{2}>0$, $\mathcal{L}^{2}>0$, $\Omega_{\varphi}^{2}>0$, $\Omega_{r}^{2}>0$ (i.e. the orbit lies outside the ISCO), $R>R_{h}$ with ($R^2=(r^2+b^2)$) and $\nu_{U}>\nu_{L}>0$; inadmissible points are given a smooth penalty. The uncertainties shown below are one-dimensional profile-likelihood intervals, $\Delta\chi^{2}=\chi^{2}-\chi^{2}_{\min}=1$, derived by re-optimizing all other parameters for each value of the parameter of interest. Each configuration is classified as a regular black hole ($b<R_{h}$) or a traversable wormhole ($b>R_{h}$), where $R_{h}$ is the biggest positive root of $R^{4}+l^{2}R^{2}-2l^{2}R +l^{2}Q^{2}=0$; for $l=100M$, a horizon exists for $Q/M\lesssim0.9995$.
\subsubsection{Degeneracies}
Read these constraints carefully; the profile analysis makes the reason explicit. The data do not constrain the regularization parameter $b$. For both sources, $\chi^{2}$ varies by less than $0.4$ over the entire range $0\le b/M\le3$, over $Q$, $r$ and $M$, and does so non-monotonically, indicating that the residual variation is of the order of the fit's numerical resolution rather than a genuine constraint. Physically, $b$ enters the frequencies only via $b^{2}/R^{2}\lesssim0.03$ at the relevant emission radii, much below the resolution of two frequency measurements per source.

The charge and the AdS curvature radius $l$ are quite degenerate, since the two terms in Eq.~\eqref{function_f} raise $\nu_{L}/\nu_{U}$ at fixed $r$ in the same way. Table~\ref{tab:l} profiles $\chi^{2}$ over $l$ at $Q=b=0$: instead of improving monotonically, the fit has a pronounced minimum near $l\simeq40M$ ($\chi^{2}=0.214$ and $0.097$ for GRO~J1655$-$40 and XTE~J1550$-$564), essentially as good as the charged fit of Eq.~\eqref{eq:jointQ}, and without any charge at all. Both $l\to20M$ (over-confinement) and $l\to1000M$ (the fit relaxing back toward the disfavored Schwarzschild value) degrade the fit. The data therefore constrain a combination of $Q$ and $l$, not either parameter independently: the value quoted in Eq.~\eqref{eq:jointQ} holds at fixed $l=100M$, and an equally acceptable description can be found with $Q=0$ at $l\simeq40M$.

Ultimately, the metric~\eqref{metric} is static, whereas both sources have independently estimated spins, and frame dragging elevates $\nu_{\varphi}$ relative to $\nu_{r}$ in the same way as $Q$ or a smaller $l$. The deformation seen here likely accounts for the overlooked rotation in addition to (or instead of) any genuine charge or AdS curvature, and should be interpreted as an effective representation of the deviation from Schwarzschild rather than as an assessment of each factor separately. Breaking this degeneracy requires a rotating generalization of the metric, or independent timing information such as a genuine low-frequency (nodal-precession) QPO, which the current static geometry cannot generate.

We conclude that physically admissible configurations of the charged Simpson-Visser-AdS spacetime replicate both observed HF-QPO pairs at high statistical confidence, that the data require a departure from the Schwarzschild limit, but that the individual parameters $b$, $Q$ and $l$ remain degenerate. Read the entries of Table~\ref{tab:qpo} as allowed configurations subject to these degeneracies, rather than as independent determinations of the spacetime parameters.
\begin{table}
\caption{\label{tab:qpo}Best fit configurations of the twin-peak HF-QPOs of GRO~J1655$-$40 and XTE~J1550$-$564 in the charged Simpson-Visser-AdS spacetime ($b=2M$ represent wormholes), for $l=100M$. The emission radius $r/M$ and the mass are free parameters, with uncertainties given by $\Delta\chi^{2}=1$ profile intervals. For three data points and two free parameters, $\nu_{\rm dof}=1$.}
\begin{ruledtabular}
\begin{tabular}{ccccccc}
$Q/M$ & $b/M$ & $r/M$ & $M/M_\odot$ & $\nu_U$ (Hz) & $\nu_L$ (Hz) & $\chi^2_{\rm min}$ \\ \hline
\multicolumn{7}{c}{GRO J1655-40} \\ \hline
0.0 & 0.0 & $6.421\pm0.045$ & $4.488\pm0.060$ & 448.4 & 301.9 & 9.68 \\
0.0 & 1.0 & $6.356\pm0.046$ & $4.475\pm0.060$ & 448.3 & 302.1 & 10.01 \\
0.0 & 2.0 & $6.162\pm0.052$ & $4.425\pm0.065$ & 448.8 & 303.2 & 11.11 \\
0.2 & 0.0 & $6.356\pm0.045$ & $4.538\pm0.064$ & 448.7 & 303.1 & 8.85 \\
0.2 & 1.0 & $6.291\pm0.046$ & $4.525\pm0.065$ & 448.5 & 303.1 & 9.16 \\
0.2 & 2.0 & $6.113\pm0.051$ & $4.463\pm0.065$ & 448.4 & 301.9 & 10.20 \\
0.4 & 0.0 & $6.194\pm0.044$ & $4.663\pm0.064$ & 449.1 & 302.0 & 6.29 \\
0.4 & 1.0 & $6.129\pm0.046$ & $4.650\pm0.064$ & 448.6 & 301.7 & 6.58 \\
0.4 & 2.0 & $5.935\pm0.052$ & $4.588\pm0.070$ & 449.2 & 302.3 & 7.63 \\
0.6 & 0.0 & $5.886\pm0.044$ & $4.938\pm0.069$ & 448.8 & 301.1 & 2.59 \\
0.6 & 1.0 & $5.822\pm0.046$ & $4.913\pm0.071$ & 448.8 & 300.9 & 2.83 \\
0.6 & 2.0 & $5.611\pm0.052$ & $4.850\pm0.074$ & 449.1 & 302.4 & 3.68 \\
0.8 & 0.0 & $5.401\pm0.045$ & $5.425\pm0.079$ & 449.5 & 299.8 & 0.04 \\
0.8 & 1.0 & $5.336\pm0.047$ & $5.375\pm0.081$ & 450.4 & 299.8 & 0.02 \\
0.8 & 2.0 & $5.126\pm0.055$ & $5.275\pm0.085$ & 450.4 & 300.2 & 0.19 \\
\multicolumn{7}{c}{XTE~J1550$-$564} \\ \hline
0.0 & 0.0 & $6.388\pm0.070$ & $7.395\pm0.163$ & 274.1 & 186.9 & 8.81 \\
0.0 & 1.0 & $6.324\pm0.073$ & $7.365\pm0.165$ & 274.4 & 187.2 & 9.06 \\
0.0 & 2.0 & $6.129\pm0.081$ & $7.305\pm0.171$ & 273.8 & 187.0 & 9.85 \\
0.2 & 0.0 & $6.340\pm0.070$ & $7.455\pm0.167$ & 274.1 & 186.4 & 8.13 \\
0.2 & 1.0 & $6.275\pm0.072$ & $7.425\pm0.168$ & 274.3 & 186.6 & 8.37 \\
0.2 & 2.0 & $6.081\pm0.081$ & $7.350\pm0.172$ & 274.1 & 186.6 & 9.17 \\
0.4 & 0.0 & $6.162\pm0.070$ & $7.695\pm0.172$ & 274.2 & 186.7 & 6.14 \\
0.4 & 1.0 & $6.097\pm0.073$ & $7.665\pm0.174$ & 274.2 & 186.6 & 6.37 \\
0.4 & 2.0 & $5.903\pm0.081$ & $7.560\pm0.179$ & 274.5 & 186.8 & 7.15 \\
0.6 & 0.0 & $5.870\pm0.070$ & $8.100\pm0.187$ & 274.6 & 185.5 & 3.07 \\
0.6 & 1.0 & $5.789\pm0.073$ & $8.085\pm0.186$ & 274.8 & 186.5 & 3.27 \\
0.6 & 2.0 & $5.595\pm0.082$ & $7.965\pm0.195$ & 274.5 & 185.8 & 3.97 \\
0.8 & 0.0 & $5.385\pm0.071$ & $8.880\pm0.209$ & 275.8 & 185.1 & 0.19 \\
0.8 & 1.0 & $5.320\pm0.073$ & $8.820\pm0.214$ & 275.6 & 184.5 & 0.25 \\
0.8 & 2.0 & $5.109\pm0.085$ & $8.655\pm0.222$ & 275.5 & 184.6 & 0.59 \\
\end{tabular}
\end{ruledtabular}
\end{table}
\begin{table}
\caption{\label{tab:2}Profile $\chi^{2}$ in the charge parameter at $l=100M$, with $b$, $r$ and $M$ re-optimized at each $Q$. $\Delta\chi^{2}=1$, $2.71$ and $6.63$ correspond to the $68\%$, $90\%$ and $99\%$ intervals for one parameter of interest.}
\begin{ruledtabular}
\begin{tabular}{ccccc}
 & \multicolumn{2}{c}{GRO~J1655$-$40} & \multicolumn{2}{c}{XTE~J1550$-$564} \\
$Q/M$ & $\chi^{2}$ & $\Delta\chi^{2}$ & $\chi^{2}$ & $\Delta\chi^{2}$ \\ \hline
0.00 & 9.719 & 9.617 & 8.875 & 8.857 \\
0.20 & 8.814 & 8.712 & 8.204 & 8.186 \\
0.40 & 6.388 & 6.287 & 6.194 & 6.177 \\
0.50 & 4.534 & 4.432 & 4.734 & 4.716 \\
0.60 & 2.587 & 2.485 & 3.150 & 3.132 \\
0.70 & 0.859 & 0.757 & 1.499 & 1.482 \\
0.75 & 0.366 & 0.264 & 0.754 & 0.736 \\
0.80 & 0.140 & 0.038 & 0.242 & 0.224 \\
0.85 & 0.102 & 0.000 & 0.018 & 0.000 \\
0.90 & 0.739 & 0.637 & 0.050 & 0.033 \\
0.95 & 2.923 & 2.821 & 0.759 & 0.742 \\
1.00 & 7.253 & 7.152 & 2.715 & 2.697 \\
\end{tabular}
\end{ruledtabular}
\end{table}
\begin{table}\caption{\label{tab:l}
The $\chi^{2}$ profile as a function of the AdS radius $l$ at $Q=b=0$. $\Delta\chi^{2}$ is measured from each source's own minimum over this grid ($l=40M$ for both). The degeneracy mentioned in the text is driven by the near-degeneracy between this minimum and the charged best fit of Eq.~\eqref{eq:jointQ}.}
\begin{ruledtabular}
\begin{tabular}{ccccc}
 & \multicolumn{2}{c}{GRO~J1655$-$40} & \multicolumn{2}{c}{XTE~J1550$-$564} \\
$l/M$ & $\chi^2$ & $\Delta\chi^2$ & $\chi^2$ & $\Delta\chi^2$ \\ \hline
20 & 37.011 & 36.797 & 19.489 & 19.390 \\
30 & 5.898 & 5.684 & 2.147 & 2.050 \\
40 & 0.214 & 0.000 & 0.097 & 0.000 \\
50 & 1.104 & 0.890 & 1.504 & 1.407 \\
70 & 5.077 & 4.862 & 5.070 & 4.973 \\
100 & 9.719 & 9.505 & 8.875 & 8.778 \\
200 & 15.540 & 15.326 & 13.065 & 12.968 \\
1000 & 17.997 & 17.783 & 14.894 & 14.797 \\
\end{tabular}
\end{ruledtabular}
\end{table}
\subsubsection{Results}
Table~\ref{tab:qpo} collects the best-fit configurations on a grid in $(Q/M,b/M)$ at fixed $l=100M$. The two features are particularly relevant: first, within the asymptotically flat Schwarzschild limit, $Q=b=0$ and $l\rightarrow\infty$, the best fit requires $r=6.696M$, $M=4.163\,M_{\odot}$ for GRO~J1655$-$40 and $r=6.647M$, $M=6.885\,M_{\odot}$ for XTE~J1550$-$564, giving $\chi^{2}=17.97$ and $14.92$, respectively, for one degree of freedom ($p\sim10^{-5}-10^{-4}$). The corresponding small $p$-values, on the order of $10^{-5}$, indicate substantial tension between the HF-QPO frequency ratio and the independently measured masses within the static Schwarzschild model. In particular, reproducing the observed frequency ratio while retaining the Schwarzschild geometry drives the fitted masses below the dynamical mass measurements.

Introducing a nonzero $Q$ at $l=100M$ steadily improves the fit for both sources (for details, please see Table~\ref{tab:qpo}), and a joint fit in which the two sources share a single value of $Q$ at $b=0$, six data points, five free parameters ($r_{1,2}$, $M_{1,2}$, $Q$), $\nu_{\rm dof}=1$, gives us
\begin{equation}
Q/M=0.810^{+0.060}_{-0.062}, \qquad  \chi^{2}_{\rm min}=0.179\, , 
\label{eq:jointQ}
\end{equation}
At this configuration, the best-fit emission radii are $r=5.40M$ and $r=5.39M$ for GRO~J1655$-$40 and XTE~J1550$-$564, respectively, while the recovered masses are $5.43\,M_{\odot}$ and $8.85$--$8.88\,M_{\odot}$. These values are consistent with the corresponding dynamical mass measurements. Both emission radii lie outside the ISCO ($r_{\rm ISCO}=4.777M$) and the horizon ($R_h=1.586M$). The configuration is also sub-extremal, since $Q/M=0.810<Q_{\rm ext}/M\simeq0.9999$ for $l=100M$.

\begin{figure}
\centering
\includegraphics[width=0.9\columnwidth]{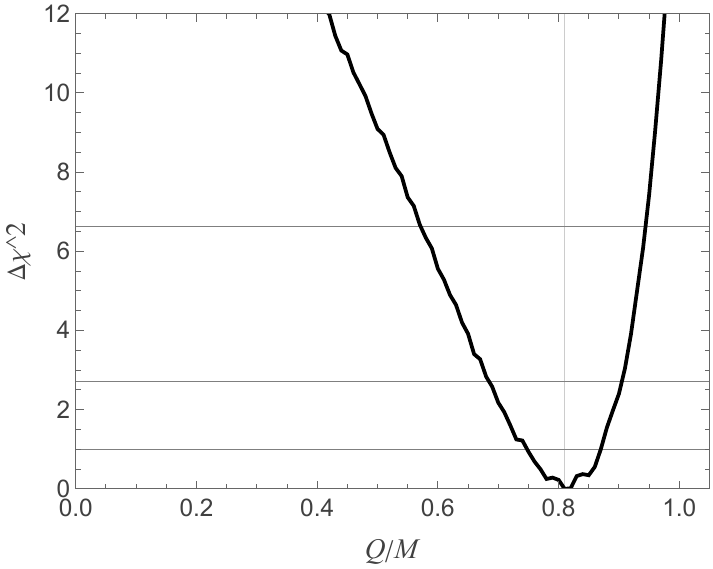}
\caption{Plot of $\Delta\chi^{2}$ as a function of $Q/M$ for GRO~J1655$-$40 and XTE~J1550$-$564 combined, at $b=0$ and $l=100M$. The horizontal lines mark $\Delta\chi^{2}=1,\,2.71,\,6.63$ (the one-parameter $68\%$, $90\%$ and $99\%$ intervals). The minimum at $Q/M=0.810^{+0.060}_{-0.062}$ excludes $Q=0$ at $4.3\sigma$.}\label{fig:dchi2}
\end{figure}
Figure~\ref{fig:dchi2} shows the joint profile of $\Delta\chi^{2}$ along $Q/M$ obtained by combining GRO~J1655$-$40 and XTE~J1550$-$564 through shared values of $Q$, at fixed $b=0$ and $l=100M$; at each value of $Q$, the emission radius and mass of each source are re-optimized independently. The curve has a well-defined, mildly asymmetric minimum at $Q/M \approx 0.810$ ($1\sigma$, $\Delta\chi^{2}=1$), with $\chi^{2}_{\rm min}=0.179$, indicating that the two sources are mutually consistent at this point. The $90\%$ and $99\%$ intervals ($\Delta\chi^{2}=2.71$ and $6.63$) widen to approximately $Q/M\simeq0.68$--$0.91$ and $0.57$--$0.94$, respectively, and the curve rises steeply toward $Q=0$: $\chi^{2}=18.59$, a $\Delta\chi^{2}=18.41$ increase over the minimum, corresponding to a $4.3\sigma$ exclusion of the uncharged configuration. The interval is visibly steeper on the low-$Q$ side than the high-$Q$ side, reflecting that $\nu_{L}/\nu_{U}$ responds more strongly to $Q$ near zero than near the sub-extremal parameter range.

The above result should not be interpreted independently of the AdS curvature scale. As shown in Table~\ref{tab:l}, $ Q=0$ provides a comparably good description at $l\simeq40M$, with individual minimum values of $\chi^2=0.214$ for GRO~J1655$-$40 and $\chi^2=0.097$ for XTE~J1550$-$564. Consequently, the apparent preference for nonzero $Q$ in Fig.~\ref{fig:dchi2} depends on the adopted value $l=100M$ and should not be interpreted as an independent detection of nonzero charge. Rather, the HF-QPO data constrain a combination of the deformation parameters $Q$ and $l$, with additional degeneracy involving $b$. The fitted value $Q/M=0.810^{+0.060}_{-0.062}$ should therefore be regarded as an effective parametrization of the departure from the Schwarzschild frequency structure within the fixed-$l=100M$ model slice, rather than as a standalone measurement of an astrophysical electric charge.
\section{Concluding remarks}\label{sec:conclusion}
In this article, we considered a charged Simpson-Visser-AdS spacetime geometry and examined circular motion, fundamental frequencies, and QPOs. Using the equation of motion, we obtain explicit expressions for the effective potential and the fundamental frequencies.
In Section~\ref{Sec:Charged-SV-AdS}, we investigate the metric function $f(r)$, phase diagram, and black hole horizons under the impact of electric charge $Q$, regularization parameter $b$, and AdS radius $l$. The roots of $f(r) = 0$ map the geometrical phase space, revealing transitions between three distinct topological classes: (I) non-extremal configurations featuring distinct inner and outer horizons, (II) extremal limits characterized by a single degenerate horizon, and (III) strictly positive, horizonless geometries. The strictly positive cases occurs at $(Q = 0.5, b = 2)$, and $(Q = 1.1, b = 0.2)$, for more details, see, Fig.~\ref{Fig:function_f}.

Examining the phase diagram, we observe that, compared with the RN-AdS black hole, the regularization parameter $b$ fundamentally restructures the regular black hole's geometry. Interestingly, $b$ shifts the critical point ($r_h = 0$) away from the origin. Since both $Q$ and $b$ supply effective repulsive contributions to the spacetime geometry, their coupled influence significantly shrinks the allowable phase space for black hole existence, driving the system toward a regular, horizonless configuration at lower parametric thresholds.

Moreover, from the horizon analysis, we noticed that the outer horizon $r_+$ initially expands rapidly with increasing $l$, a direct geometric consequence of weakening the negative cosmological constant and relaxing the background confinement. However, the repulsive contribution of electromagnetic charge $Q$ to the effective potential contracts $r_+$ and expands $r_-$. This convergence restricts the physical horizon separation, suppresses the black hole surface area, and drives the spacetime toward an extremal state ($r_+ \to r_-$).
Including $b$ introduces a highly non-linear geometric coupling to the charge. In the weak-charge regime, $b$ reinforces the effective central repulsion, working together with $Q$ to systematically reduce the horizon radii (for details, see Fig.~\ref{fig:horizons}). 

To explore particle dynamics and the stability of circular orbits, we examine the effective potential, specific angular momentum, specific energy, and the ISCO. 
From the effective potential, we observed that the Schwarzschild geometry provides a standard potential well for stable orbits, while the charged Simpson–Visser–AdS spacetime fundamentally restructures the strong-field regime. In particular, $Q$ provides an effective repulsion that raises the centrifugal barrier; this amplifies the instability maxima associated with unstable circular orbits and shifts the stable orbital domains outward. As a result, incoming particles from infinity require more energy to climb the effective potential when $Q$ and $l$ are large. Moreover, the AdS radius $l$ exerts a competing influence, systematically weakening the background curvature. 

In addition, by examining particle angular momentum and energy, we find that $Q$ and $l$ monotonically shift the ISCO inward to smaller radii while simultaneously lowering the curves. The regularization parameter $b$ has a similar effect on the kinematic thresholds, reducing the required $\mathcal{E}$ and $\mathcal{L}$ for orbital stability, though it induces only marginal shifts in the ISCO radial coordinate. Physically, this behavior indicates that the regularized core's modified spacetime geometry softens the effective gravitational potential in the strong-field regime.

Our results show a monotonic decrease of $r_{\rm ISCO}$ as both $Q$ and $b$ increase. Dynamically, this inward shift occurs because the repulsive contribution of the electromagnetic charge to the effective potential, along with $b$ modifying the central geometry, effectively softens the strong-field gravitational gradient near the core.  
Moreover, $l$ has a distinct, highly nonlinear scaling effect on $r_{\rm ISCO}$. In the deep AdS regime (small $l$), the strong confining effect of the negative cosmological constant couples heavily to the orbital dynamics, driving pronounced shifts in $r_{\rm ISCO}$. However, as $l$ increases, the spacetime smoothly dilutes toward asymptotic flatness ($\Lambda \to 0$). In this regime, the cosmological confinement rapidly decays, causing the ISCO radius to decouple from $l$ and asymptotically converge to the flat-space limit characteristic of a charged Simpson–Visser black hole.

To better understand particle dynamics, we investigated the Keplerian, radial, and vertical epicyclic frequencies, which characterize small oscillations near stable circular orbits. As expected from standard orbital kinematics, the Keplerian frequency decays monotonically with radial distance, supporting previous findings~\cite{Khan2026EPJC...86..597K}. However, in the strong-field regime, both $Q$ and $b$ introduce effective repulsive corrections to the central gravitational well. This repulsion lowers the required orbital velocity, systematically suppressing $\Omega_\phi$ in the charged Simpson–Visser–AdS spacetime relative to the singular Schwarzschild-AdS black hole. While the modifications from $Q$ and $b$ dominate the inner geometry, the AdS radius $l$ becomes increasingly influential at larger radii, governing the deviation from asymptotic flatness and further suppressing the Keplerian frequency in the outer domains.

Similarly, the regularization parameter $b$ and AdS radius $l$ decrease the radial frequency profile along $r$. Unlike the Keplerian frequency, the black hole charge $Q$ increases the radial frequency. The curves exhibit a well-defined maximum, whose position marks the transition region between the strongly curved inner geometry and the weak-field regime.
We found that $Q$ uniformly shifts the entire profile upward, increases the peak value of $\Omega_r$, and slightly displaces the maximum inward toward the black hole. As a result, the Simpson-Visser-AdS black hole has a smaller $\Omega_r$ profile than Schwarzschild-AdS and charged Simpson-Visser-AdS black holes.

Our examination of the QPOs shows that $b$ and $l$ produce different twin-peak QPOs, as the presence of $b$ yields higher QPO frequencies than in the Schwarzschild black hole. Moreover, we observe a non-linear resonance shift when $Q$ and $l$ are present simultaneously. For a given $b$, the $3:2$ resonance ratio occurs at a different radial coordinate $r$, unlike the uncharged Simpson-Visser case. 
Unlike previous investigations of black holes in GR, the electromagnetic charge, Simpson-Visser coupling, and AdS radius $l$ produce an updated ISCO. This suggests a potential observational signature: a specific range of QPO frequencies that can occur only when the vector field $\phi_\mu$, the regularization parameter $b$, the charge $Q$, and the AdS radius $l$ are non-zero. This may allow future X-ray timing missions to distinguish between geometric and field-theoretical deviations from Einstein gravity.

Within the RP framework, the charged Simpson-Visser-AdS spacetime reproduces the twin-peak HF-QPO frequencies and dynamical masses of GRO~J1655$-$40 and XTE~J1550$-$564 substantially better than the Schwarzschild limit, which is excluded at $\sim\!4\sigma$. However, this preference does not isolate a genuine astrophysical charge: at fixed AdS radius $l=100M$, the joint fit favors $Q/M\simeq0.810$, but $b$ remains entirely unconstrained by the data, and $Q$ is strongly degenerate with $l$ itself, so the fitted value is better read as an effective measure of the spacetime's overall departure from Schwarzschild than as an independent determination of any single parameter. Since the metric is static, this inferred deformation likely also absorbs the spin neglected here. Disentangling charge, AdS curvature, $b$, and spin will require either a rotating generalization of the spacetime or additional timing information—such as a genuine low-frequency nodal-precession signal beyond the two HF-QPO peaks considered in this work.
\section*{Acknowledgment}
J.R. thanks Grant No. FA-F-2021-510 of the Uzbekistan Agency for Innovative Development and the Silesian University of Opava for hospitality. The work of Weiwei Wang was supported by the NSF of Fujian Province of China (Grant Nos. 2024J011011 and 2022J01105).\\
\subsection*{Conflicts of Interest}
The authors declare no conflicts of interest.
\def\prc{Phys. Rev. C}
\def\pre{Phys. Rev. E}
\def\prd{Phys. Rev. D}
\def\prl{Physical Review Letters}
\def\jcap{Journal of Cosmology and Astroparticle Physics}
\def\apss{Astrophysics and Space Science}
\def\mnras{Monthly Notices of the Royal Astronomical Society}
\def\apj{The Astrophysical Journal}
\def\aap{Astronomy and Astrophysics}
\def\actaa{Acta Astronomica}
\def\pasj{Publications of the Astronomical Society of Japan}
\def\apjl{Astrophysical Journal Letters}
\def\pasa{Publications Astronomical Society of Australia}
\def\nat{Nature}
\def\physrep{Physics Reports}
\def\araa{Annual Review of Astronomy and Astrophysics}
\def\apjs{The Astrophysical Journal Supplement}
\def\aapr{The Astronomy and Astrophysics Review}
\def\procspie{Proceedings of the SPIE}
\bibliographystyle{apsrev4-1}
\bibliography{References}

@ARTICLE{Khan2026EPJC...86..597K,
       author = {{Khan}, Saeed Ullah and {Rayimbaev}, Javlon and {Hayat}, Haidar and {Chen}, Zhi-Min and {Abdullaev}, Mardon and {Murodov}, Sardor and {Wang}, Weiwei},
        title = "{Quasi-periodic oscillations in rotating Simpson─Visser black holes in STVG}",
      journal = {European Physical Journal C},
         year = 2026,
        month = jun,
       volume = {86},
       number = {6},
          eid = {597},
        pages = {597},
          doi = {10.1140/epjc/s10052-026-15830-w},
       adsurl = {https://ui.adsabs.harvard.edu/abs/2026EPJC...86..597K}
}

@ARTICLE{Khan2026ChJPh.102..711K,
       author = {{Khan}, Saeed Ullah and {Zahid}, Muhammad and {Chen}, Zhi-Min and {Turaev}, Yunus and {Akhmedov}, Munisbek and {Usanov}, Sulton and {Rayimbaev}, Javlon and {Wang}, Weiwei},
        title = "{Rotating black-bounce black holes in modified gravity: Photon orbits and shadow}",
      journal = {Chinese Journal of Physics},
         year = 2026,
        month = aug,
       volume = {102},
        pages = {711-727},
          doi = {10.1016/j.cjph.2026.03.023},
       adsurl = {https://ui.adsabs.harvard.edu/abs/2026ChJPh.102..711K}
}

@ARTICLE{Simpson2019JCAP...02..042S,
       author = {{Simpson}, Alex and {Visser}, Matt},
        title = "{Black-bounce to traversable wormhole}",
      journal = {\jcap},
         year = 2019,
        month = feb,
       volume = {2019},
       number = {2},
          eid = {042},
        pages = {042},
          doi = {10.1088/1475-7516/2019/02/042},
archivePrefix = {arXiv},
       eprint = {1812.07114},
 primaryClass = {gr-qc},
       adsurl = {https://ui.adsabs.harvard.edu/abs/2019JCAP...02..042S}
}

@ARTICLE{2021JCAP...04..082M,
       author = {{Mazza}, Jacopo and {Franzin}, Edgardo and {Liberati}, Stefano},
        title = "{A novel family of rotating black hole mimickers}",
      journal = {\jcap},
         year = 2021,
        month = apr,
       volume = {2021},
       number = {4},
          eid = {082},
        pages = {082},
          doi = {10.1088/1475-7516/2021/04/082},
archivePrefix = {arXiv},
       eprint = {2102.01105},
 primaryClass = {gr-qc},
       adsurl = {https://ui.adsabs.harvard.edu/abs/2021JCAP...04..082M}
}

@ARTICLE{2021PhRvL.126j1102B,
       author = {{Bl{\'a}zquez-Salcedo}, Jose Luis and {Knoll}, Christian and {Radu}, Eugen},
        title = "{Traversable Wormholes in Einstein-Dirac-Maxwell Theory}",
      journal = {\prl},
         year = 2021,
        month = mar,
       volume = {126},
       number = {10},
          eid = {101102},
        pages = {101102},
          doi = {10.1103/PhysRevLett.126.101102},
archivePrefix = {arXiv},
       eprint = {2010.07317},
 primaryClass = {gr-qc},
       adsurl = {https://ui.adsabs.harvard.edu/abs/2021PhRvL.126j1102B}
}

@ARTICLE{2022EPJC...82..533B,
       author = {{Bl{\'a}zquez-Salcedo}, Jose Luis and {Knoll}, Christian and {Radu}, E.},
        title = "{Einstein-Dirac-Maxwell wormholes: ansatz, construction and properties of symmetric solutions}",
      journal = {European Physical Journal C},
         year = 2022,
        month = jun,
       volume = {82},
       number = {6},
          eid = {533},
        pages = {533},
          doi = {10.1140/epjc/s10052-022-10488-6},
archivePrefix = {arXiv},
       eprint = {2108.12187},
 primaryClass = {gr-qc},
       adsurl = {https://ui.adsabs.harvard.edu/abs/2022EPJC...82..533B}
}

@ARTICLE{2022PhRvL.128i1104K,
       author = {{Konoplya}, R.~A. and {Zhidenko}, A.},
        title = "{Traversable Wormholes in General Relativity}",
      journal = {\prl},
         year = 2022,
        month = mar,
       volume = {128},
       number = {9},
          eid = {091104},
        pages = {091104},
          doi = {10.1103/PhysRevLett.128.091104},
archivePrefix = {arXiv},
       eprint = {2106.05034},
 primaryClass = {gr-qc},
       adsurl = {https://ui.adsabs.harvard.edu/abs/2022PhRvL.128i1104K}
}

@ARTICLE{2023EPJC...83..854V,
       author = {{Vrba}, Jaroslav and {Rayimbaev}, Javlon and {Stuchlik}, Zdenek and {Ahmedov}, Bobomurat},
        title = "{Charged particles motion and quasiperiodic oscillation in Simpson─Visser spacetime in the presence of external magnetic fields}",
      journal = {European Physical Journal C},
         year = 2023,
        month = sep,
       volume = {83},
       number = {9},
          eid = {854},
        pages = {854},
          doi = {10.1140/epjc/s10052-023-12023-7},
       adsurl = {https://ui.adsabs.harvard.edu/abs/2023EPJC...83..854V}
}

@ARTICLE{Ahmed2026EPJC...86..658A,
       author = {{Ahmed}, Faizuddin and {Al-Badawi}, Ahmad and {Fathi}, Mohsen},
        title = "{Charged Simpson─Visser AdS black holes: geodesic structure and thermodynamic properties}",
      journal = {European Physical Journal C},
         year = 2026,
        month = jun,
       volume = {86},
       number = {6},
          eid = {658},
        pages = {658},
          doi = {10.1140/epjc/s10052-026-15935-2},
archivePrefix = {arXiv},
       eprint = {2601.10469},
 primaryClass = {gr-qc},
       adsurl = {https://ui.adsabs.harvard.edu/abs/2026EPJC...86..658A}
}

@ARTICLE{2025PhRvD.112l4018D,
       author = {{Dasgupta}, Anirban and {Banerjee}, Indrani},
        title = "{Constraining the rotating Simpson-Visser spacetime from the observed quasiperiodic oscillations in black holes}",
      journal = {\prd},
         year = 2025,
        month = dec,
       volume = {112},
       number = {12},
          eid = {124018},
        pages = {124018},
          doi = {10.1103/2x2n-t383},
archivePrefix = {arXiv},
       eprint = {2509.15761},
 primaryClass = {gr-qc},
       adsurl = {https://ui.adsabs.harvard.edu/abs/2025PhRvD.112l4018D}
}

@ARTICLE{Ingram2010MNRAS,
       author = {{Ingram}, Adam and {Done}, Chris},
        title = "{A physical interpretation of the variability power spectral components in accreting neutron stars}",
      journal = {Mon.Not.R.Astron.Soc},
         year = 2010,
        month = jul,
       volume = {405},
       number = {4},
        pages = {2447-2452},
          doi = {10.1111/j.1365-2966.2010.16614.x},
archivePrefix = {arXiv},
       eprint = {0907.5485},
 primaryClass = {astro-ph.SR},
       adsurl = {https://ui.adsabs.harvard.edu/abs/2010MNRAS.405.2447I}
}

@BOOK{Chandrasekhar1983mtbh.book.C,
       author = {{Chandrasekhar}, S.},
        title = "{The mathematical theory of black holes}",
         year = 1983,
       adsurl = {https://ui.adsabs.harvard.edu/abs/1983mtbh.book.....C}
}

@BOOK{Bambi2017bhlt.book..B,
       author = {{Bambi}, Cosimo},
        title = "{Black Holes: A Laboratory for Testing Strong Gravity}",
         year = 2017,
          doi = {10.1007/978-981-10-4524-0},
       adsurl = {https://ui.adsabs.harvard.edu/abs/2017bhlt.book.....B}
}

@ARTICLE{Zdunik2000AA,
       author = {{Zdunik}, J.~L. and {Haensel}, P. and {Gondek-Rosi{\'n}ska}, D. and {Gourgoulhon}, E.},
        title = "{Innermost stable circular orbits around strange stars and kHz QPOs in low-mass X-ray binaries}",
      journal = {Astron.Astrophys.},
         year = 2000,
        month = apr,
       volume = {356},
        pages = {612-618},
archivePrefix = {arXiv},
       eprint = {astro-ph/0002394},
 primaryClass = {astro-ph},
       adsurl = {https://ui.adsabs.harvard.edu/abs/2000A&A...356..612Z}
}

@ARTICLE{Wilkins2012MNRAS,
       author = {{Wilkins}, D.~R. and {Fabian}, A.~C.},
        title = "{Understanding X-ray reflection emissivity profiles in AGN: locating the X-ray source}",
      journal = {\mnras},
         year = 2012,
        month = aug,
       volume = {424},
       number = {2},
        pages = {1284-1296},
          doi = {10.1111/j.1365-2966.2012.21308.x},
archivePrefix = {arXiv},
       eprint = {1205.3179},
 primaryClass = {astro-ph.HE},
       adsurl = {https://ui.adsabs.harvard.edu/abs/2012MNRAS.424.1284W}
}

@ARTICLE{Aliev2002MNRAS.336..241A,
       author = {{Aliev}, A.~N. and {{\"O}zdemir}, N.},
        title = "{Motion of charged particles around a rotating black hole in a magnetic field}",
      journal = {\mnras},
         year = 2002,
        month = oct,
       volume = {336},
       number = {1},
        pages = {241-248},
          doi = {10.1046/j.1365-8711.2002.05727.x},
archivePrefix = {arXiv},
       eprint = {gr-qc/0208025},
 primaryClass = {gr-qc},
       adsurl = {https://ui.adsabs.harvard.edu/abs/2002MNRAS.336..241A}
}

@ARTICLE{Vrba2020PhRvD.101l4039V,
       author = {{Vrba}, Jaroslav and {Abdujabbarov}, Ahmadjon and {Kolo{\v{s}}}, Martin and {Ahmedov}, Bobomurat and {Stuchl{\'\i}k}, Zden{\v{e}}k and {Rayimbaev}, Javlon},
        title = "{Charged and magnetized particles motion in the field of generic singular black holes governed by general relativity coupled to nonlinear electrodynamics}",
      journal = {\prd},
         year = 2020,
        month = jun,
       volume = {101},
       number = {12},
          eid = {124039},
        pages = {124039},
          doi = {10.1103/PhysRevD.101.124039},
       adsurl = {https://ui.adsabs.harvard.edu/abs/2020PhRvD.101l4039V}
}

@ARTICLE{Zahid2022EPJC...82..494Z,
       author = {{Zahid}, Muhammad and {Rayimbaev}, Javlon and {Khan}, Saeed Ullah and {Ren}, Jingli and {Ahmedov}, Saidmuhammad and {Ibragimov}, Inomjon},
        title = "{Dynamics and collisions of magnetized particles around charged black holes in Einstein-Maxwell-scalar theory}",
      journal = {European Physical Journal C},
         year = 2022,
        month = may,
       volume = {82},
       number = {5},
          eid = {494},
        pages = {494},
          doi = {10.1140/epjc/s10052-022-10432-8},
       adsurl = {https://ui.adsabs.harvard.edu/abs/2022EPJC...82..494Z}
}

@ARTICLE{Amendola2008JCAP...04..013A,
       author = {{Amendola}, Luca and {Kunz}, Martin and {Sapone}, Domenico},
        title = "{Measuring the dark side (with weak lensing)}",
      journal = {\jcap},
         year = 2008,
        month = apr,
       volume = {2008},
       number = {4},
          eid = {013},
        pages = {013},
          doi = {10.1088/1475-7516/2008/04/013},
archivePrefix = {arXiv},
       eprint = {0704.2421},
 primaryClass = {astro-ph},
       adsurl = {https://ui.adsabs.harvard.edu/abs/2008JCAP...04..013A}
}

@ARTICLE{2025EPJC...85.1029N,
       author = {{Nishonov}, Isomiddin and {Murodov}, Sardor and {Ahmedov}, Bobomurat and {Khan}, Saeed Ullah and {Rayimbaev}, Javlon and {Ibragimov}, Inomjon and {Sabirov}, Sardor},
        title = "{QPOs from charged particles around charged black holes in STVG}",
      journal = {European Physical Journal C},
         year = 2025,
        month = sep,
       volume = {85},
       number = {9},
          eid = {1029},
        pages = {1029},
          doi = {10.1140/epjc/s10052-025-14751-4},
       adsurl = {https://ui.adsabs.harvard.edu/abs/2025EPJC...85.1029N}
}

@ARTICLE{2025PDU....5002110S,
       author = {{Shermatov}, Abubakir and {Rayimbaev}, Javlon and {Murodov}, Sardor and {L{\"u}tf{\"u}o{\u{g}}lu}, Bekir Can and {Ahmedov}, Bobomurat and {Zahid}, Muhammad and {Ibragimov}, Inomjon and {Shermatov}, Bahran},
        title = "{Phantom black holes in f(R,T) gravity: From circular orbits to QPO tests}",
      journal = {Physics of the Dark Universe},
         year = 2025,
        month = dec,
       volume = {50},
          eid = {102110},
        pages = {102110},
          doi = {10.1016/j.dark.2025.102110},
       adsurl = {https://ui.adsabs.harvard.edu/abs/2025PDU....5002110S}
}

@ARTICLE{2024EPJC...84.1114R,
       author = {{Rayimbaev}, Javlon and {Murodov}, Sardor and {Shermatov}, Abubakir and {Yusupov}, Amirkhon},
        title = "{QPOs from charged particles around magnetized black holes in braneworlds}",
      journal = {European Physical Journal C},
         year = 2024,
        month = oct,
       volume = {84},
       number = {10},
          eid = {1114},
        pages = {1114},
          doi = {10.1140/epjc/s10052-024-13463-5},
       adsurl = {https://ui.adsabs.harvard.edu/abs/2024EPJC...84.1114R}
}

@ARTICLE{2021EPJC...81.1067S,
       author = {{Shahzadi}, Misbah and {Kolo{\v{s}}}, Martin and {Stuchl{\'\i}k}, Zden{\v{e}}k and {Habib}, Yousaf},
        title = "{Epicyclic oscillations in spinning particle motion around Kerr black hole applied in models fitting the quasi-periodic oscillations observed in microquasars and AGNs}",
      journal = {European Physical Journal C},
         year = 2021,
        month = dec,
       volume = {81},
       number = {12},
          eid = {1067},
        pages = {1067},
          doi = {10.1140/epjc/s10052-021-09868-1},
archivePrefix = {arXiv},
       eprint = {2104.09640},
 primaryClass = {astro-ph.HE},
       adsurl = {https://ui.adsabs.harvard.edu/abs/2021EPJC...81.1067S}
}

@ARTICLE{Stella1999ApJ,
       author = {{Stella}, Luigi and {Vietri}, Mario and {Morsink}, Sharon M.},
        title = "{Correlations in the Quasi-periodic Oscillation Frequencies of Low-Mass X-Ray Binaries and the Relativistic Precession Model}",
      journal = {The Astrophysical Journal},
         year = 1999,
        month = oct,
       volume = {524},
       number = {1},
        pages = {L63-L66},
          doi = {10.1086/312291},
archivePrefix = {arXiv},
       eprint = {astro-ph/9907346},
 primaryClass = {astro-ph},
       adsurl = {https://ui.adsabs.harvard.edu/abs/1999ApJ...524L..63S}
}

@ARTICLE{EHT2019ApJ...875L...1E,
       author = {{Event Horizon Telescope Collaboration}},
        title = "{First M87 Event Horizon Telescope Results. I. The Shadow of the Supermassive Black Hole}",
      journal = {\apjl},
         year = 2019,
        month = apr,
       volume = {875},
       number = {1},
          eid = {L1},
        pages = {L1},
          doi = {10.3847/2041-8213/ab0ec7},
archivePrefix = {arXiv},
       eprint = {1906.11238},
 primaryClass = {astro-ph.GA},
       adsurl = {https://ui.adsabs.harvard.edu/abs/2019ApJ...875L...1E}
}

@ARTICLE{EHT2023ApJ...957L..20E,
       author = {{Event Horizon Telescope Collaboration} },
        title = "{First M87 Event Horizon Telescope Results. IX. Detection of Near-horizon Circular Polarization}",
      journal = {\apjl},
         year = 2023,
        month = nov,
       volume = {957},
       number = {2},
          eid = {L20},
        pages = {L20},
          doi = {10.3847/2041-8213/acff70},
archivePrefix = {arXiv},
       eprint = {2311.10976},
 primaryClass = {astro-ph.HE},
       adsurl = {https://ui.adsabs.harvard.edu/abs/2023ApJ...957L..20E}
}

@ARTICLE{EHT2022ApJ...930L..17E,
       author = {{Event Horizon Telescope Collaboration}},
        title = "{First Sagittarius A* Event Horizon Telescope Results. VI. Testing the Black Hole Metric}",
      journal = {\apjl},
         year = 2022,
        month = may,
       volume = {930},
       number = {2},
          eid = {L17},
        pages = {L17},
          doi = {10.3847/2041-8213/ac6756},
archivePrefix = {arXiv},
       eprint = {2311.09484},
 primaryClass = {astro-ph.HE},
       adsurl = {https://ui.adsabs.harvard.edu/abs/2022ApJ...930L..17E}
}

@ARTICLE{EHT2022ApJ...930L..12E,
       author = {{Event Horizon Telescope Collaboration}},
        title = "{First Sagittarius A* Event Horizon Telescope Results. I. The Shadow of the Supermassive Black Hole in the Center of the Milky Way}",
      journal = {\apjl},
         year = 2022,
        month = may,
       volume = {930},
       number = {2},
          eid = {L12},
        pages = {L12},
          doi = {10.3847/2041-8213/ac6674},
archivePrefix = {arXiv},
       eprint = {2311.08680},
 primaryClass = {astro-ph.HE},
       adsurl = {https://ui.adsabs.harvard.edu/abs/2022ApJ...930L..12E}
}

@article{khan2024circular,
  title={Circular motion and collisions of particles with magnetic dipole moment and electric charge in dipolar magnetosphere around Schwarzschild black holes},
  author={Khan, Saeed Ullah and Abdurkhmonov, Ozodbek and Rayimbaev, Javlon and Ahmedov, Saidmuhammad and Turaev, Yunus and Muminov, Sokhibjan},
  journal={The European Physical Journal C},
  volume={84},
  number={6},
  pages={650},
  year={2024},
  publisher={Springer}
}

@ARTICLE{2022ChJPh..78..141K,
       author = {{Khan}, Saeed Ullah and {Ren}, Jingli},
        title = "{Geodesics and optical properties of a rotating black hole in Randall-Sundrum brane with a cosmological constant}",
      journal = {Chinese Journal of Physics},
         year = 2022,
        month = aug,
       volume = {78},
        pages = {141-154},
          doi = {10.1016/j.cjph.2022.06.017},
       adsurl = {https://ui.adsabs.harvard.edu/abs/2022ChJPh..78..141K}
}

@ARTICLE{2024EPJC...84..964J,
       author = {{Jumaniyozov}, Shokhzod and {Khan}, Saeed Ullah and {Rayimbaev}, Javlon and {Abdujabbarov}, Ahmadjon and {Urinbaev}, Sharofiddin and {Murodov}, Sardor},
        title = "{Circular motion and QPOs near black holes in Kalb{\textendash}Ramond gravity}",
      journal = {European Physical Journal C},
         year = 2024,
        month = sep,
       volume = {84},
       number = {9},
          eid = {964},
        pages = {964},
          doi = {10.1140/epjc/s10052-024-13351-y},
       adsurl = {https://ui.adsabs.harvard.edu/abs/2024EPJC...84..964J}
}

@ARTICLE{Pitkin2011LRR.14..5P,
       author = {{Pitkin}, Matthew and {Reid}, Stuart and {Rowan}, Sheila and {Hough}, James},
        title = "{Gravitational Wave Detection by Interferometry (Ground and Space)}",
      journal = {Living Reviews in Relativity},
         year = 2011,
        month = jul,
       volume = {14},
       number = {1},
          eid = {5},
        pages = {5},
          doi = {10.12942/lrr-2011-5},
archivePrefix = {arXiv},
       eprint = {1102.3355},
 primaryClass = {astro-ph.IM},
       adsurl = {https://ui.adsabs.harvard.edu/abs/2011LRR....14....5P}
}

@ARTICLE{Liu2024PhRvD.109h4074L,
       author = {{Liu}, Tonghua and {Biesiada}, Marek and {Tian}, Shuxun and {Liao}, Kai},
        title = "{Robust test of general relativity at the galactic scales by combining strong lensing systems and gravitational wave standard sirens}",
      journal = {\prd},
         year = 2024,
        month = apr,
       volume = {109},
       number = {8},
          eid = {084074},
        pages = {084074},
          doi = {10.1103/PhysRevD.109.084074},
archivePrefix = {arXiv},
       eprint = {2404.05907},
 primaryClass = {gr-qc},
       adsurl = {https://ui.adsabs.harvard.edu/abs/2024PhRvD.109h4074L}
}

@ARTICLE{Li2026arXiv260214458L,
       author = {{Li}, Ziyang and {Liu}, Shou-Qi and {Huang}, Jia-Hui},
        title = "{Circular orbits and observational features of the rotating Simpson-Visser black hole surrounded by a thin accretion disk}",
      journal = {arXiv e-prints},
         year = 2026,
        month = feb,
          eid = {arXiv:2602.14458},
        pages = {arXiv:2602.14458},
          doi = {10.48550/arXiv.2602.14458},
archivePrefix = {arXiv},
       eprint = {2602.14458},
 primaryClass = {astro-ph.HE},
       adsurl = {https://ui.adsabs.harvard.edu/abs/2026arXiv260214458L}
}

@ARTICLE{Strohmayer_2001ApJ...552L..49S,
       author = {{Strohmayer}, Tod E.},
        title = "{Discovery of a 450 HZ Quasi-periodic Oscillation from the Microquasar GRO J1655-40 with the Rossi X-Ray Timing Explorer}",
      journal = {\apjl},
         year = 2001,
        month = may,
       volume = {552},
       number = {1},
        pages = {L49-L53},
          doi = {10.1086/320258},
       adsurl = {https://ui.adsabs.harvard.edu/abs/2001ApJ...552L..49S}
}

@ARTICLE{2015EPJC...75..162B,
       author = {{Bambi}, Cosimo},
        title = "{Testing the nature of the black hole candidate in GRO J1655-40 with the relativistic precession model}",
      journal = {European Physical Journal C},
         year = 2015,
        month = apr,
       volume = {75},
          eid = {162},
        pages = {162},
          doi = {10.1140/epjc/s10052-015-3396-7},
archivePrefix = {arXiv},
       eprint = {1312.2228},
 primaryClass = {gr-qc},
       adsurl = {https://ui.adsabs.harvard.edu/abs/2015EPJC...75..162B}
}

@ARTICLE{Beer2002MNRAS.331..351B,
       author = {{Beer}, Martin E. and {Podsiadlowski}, Philipp},
        title = "{The quiescent light curve and the evolutionary state of GRO J1655-40}",
      journal = {\mnras},
         year = 2002,
        month = mar,
       volume = {331},
       number = {2},
        pages = {351-360},
          doi = {10.1046/j.1365-8711.2002.05189.x},
archivePrefix = {arXiv},
       eprint = {astro-ph/0109136},
 primaryClass = {astro-ph},
       adsurl = {https://ui.adsabs.harvard.edu/abs/2002MNRAS.331..351B}
}

@ARTICLE{Orosz2011ApJ...742...84O,
       author = {{Orosz}, Jerome A. and {McClintock}, Jeffrey E. and {Aufdenberg}, Jason P. and {Remillard}, Ronald A. and {Reid}, Mark J. and {Narayan}, Ramesh and {Gou}, Lijun},
        title = "{The Mass of the Black Hole in Cygnus X-1}",
      journal = {\apj},
         year = 2011,
        month = dec,
       volume = {742},
       number = {2},
          eid = {84},
        pages = {84},
          doi = {10.1088/0004-637X/742/2/84},
archivePrefix = {arXiv},
       eprint = {1106.3689},
 primaryClass = {astro-ph.HE},
       adsurl = {https://ui.adsabs.harvard.edu/abs/2011ApJ...742...84O}
}

@ARTICLE{2023MNRAS.523..375S,
       author = {{Shaikh}, Rajibul},
        title = "{Testing black hole mimickers with the Event Horizon Telescope image of Sagittarius A*}",
      journal = {\mnras},
         year = 2023,
        month = jul,
       volume = {523},
       number = {1},
        pages = {375-384},
          doi = {10.1093/mnras/stad1383},
archivePrefix = {arXiv},
       eprint = {2208.01995},
 primaryClass = {gr-qc},
       adsurl = {https://ui.adsabs.harvard.edu/abs/2023MNRAS.523..375S}
}

@ARTICLE{Rezzolla2014PhRvD,
       author = {{Rezzolla}, Luciano and {Zhidenko}, Alexander},
        title = "{New parametrization for spherically symmetric black holes in metric theories of gravity}",
      journal = {\prd},
         year = 2014,
        month = oct,
       volume = {90},
       number = {8},
          eid = {084009},
        pages = {084009},
          doi = {10.1103/PhysRevD.90.084009},
archivePrefix = {arXiv},
       eprint = {1407.3086},
 primaryClass = {gr-qc},
       adsurl = {https://ui.adsabs.harvard.edu/abs/2014PhRvD..90h4009R}
}

@ARTICLE{Konoplya2020PhRvD.101l4004K,
       author = {{Konoplya}, R.~A. and {Zhidenko}, A.},
        title = "{General parametrization of black holes: The only parameters that matter}",
      journal = {\prd},
         year = 2020,
        month = jun,
       volume = {101},
       number = {12},
          eid = {124004},
        pages = {124004},
          doi = {10.1103/PhysRevD.101.124004},
archivePrefix = {arXiv},
       eprint = {2001.06100},
 primaryClass = {gr-qc},
       adsurl = {https://ui.adsabs.harvard.edu/abs/2020PhRvD.101l4004K}
}

@ARTICLE{Rayimbaev2021Galax.9.75R,
       author = {{Rayimbaev}, Javlon and {Tadjimuratov}, Pulat and {Abdujabbarov}, Ahmadjon and {Ahmedov}, Bobomurat and {Khudoyberdieva}, Malika},
        title = "{Dynamics of Test Particles and Twin Peaks QPOs around Regular Black Holes in Modified Gravity}",
      journal = {Galaxies},
         year = 2021,
        month = oct,
       volume = {9},
       number = {4},
          eid = {75},
        pages = {75},
          doi = {10.3390/galaxies9040075},
archivePrefix = {arXiv},
       eprint = {2010.12863},
 primaryClass = {gr-qc},
       adsurl = {https://ui.adsabs.harvard.edu/abs/2021Galax...9...75R}
}

@ARTICLE{Rayimbaev2023EPJC...83..572R,
       author = {{Rayimbaev}, Javlon and {Dialektopoulos}, Konstantinos F. and {Sarikulov}, Furkat and {Abdujabbarov}, Ahmadjon},
        title = "{Quasiperiodic oscillations around hairy black holes in Horndeski gravity}",
      journal = {European Physical Journal C},
         year = 2023,
        month = jul,
       volume = {83},
       number = {7},
          eid = {572},
        pages = {572},
          doi = {10.1140/epjc/s10052-023-11769-4},
archivePrefix = {arXiv},
       eprint = {2307.03019},
 primaryClass = {gr-qc},
       adsurl = {https://ui.adsabs.harvard.edu/abs/2023EPJC...83..572R}
}

@ARTICLE{Jumaniyozov2024EPJC964y,
       author = {{Jumaniyozov}, Shokhzod and {Khan}, Saeed Ullah and {Rayimbaev}, Javlon and {Abdujabbarov}, Ahmadjon and  {Urinbaev}, Sharofiddin and {Murodov}, Sardor}, 
        title = "{Circular motion and QPOs near black holes in Kalb–Ramond gravity}",
      journal = {European Physical Journal C},
         year = 2024,
        month = mar,
       volume = {84},
       number = {964},
        doi = {10.1140/epjc/s10052-024-13351-y}
}

@ARTICLE{2026EPJC...86..510S,
       author = {{Sui}, Tao-Tao and {Long}, Chen and {Zhang}, Ye},
        title = "{The properties and predictions of quasi-periodic oscillations around a black hole in nonlocal gravity}",
      journal = {European Physical Journal C},
         year = 2026,
        month = may,
       volume = {86},
       number = {5},
          eid = {510},
        pages = {510},
          doi = {10.1140/epjc/s10052-026-15770-5},
       adsurl = {https://ui.adsabs.harvard.edu/abs/2026EPJC...86..510S}
}

@ARTICLE{2022PDU....3500930R,
       author = {{Rayimbaev}, Javlon and {Majeed}, Bushra and {Jamil}, Mubasher and {Jusufi}, Kimet and {Wang}, Anzhong},
        title = "{Quasiperiodic oscillations, quasinormal modes and shadows of Bardeen-Kiselev Black Holes}",
      journal = {Physics of the Dark Universe},
         year = 2022,
        month = mar,
       volume = {35},
          eid = {100930},
        pages = {100930},
          doi = {10.1016/j.dark.2021.100930},
archivePrefix = {arXiv},
       eprint = {2202.11509},
 primaryClass = {gr-qc},
       adsurl = {https://ui.adsabs.harvard.edu/abs/2022PDU....3500930R}
}

@ARTICLE{2023Galax..11...95R,
       author = {{Rayimbaev}, Javlon and {Abdulxamidov}, Farrux and {Tojiev}, Sardor and {Abdujabbarov}, Ahmadjon and {Holmurodov}, Farhod},
        title = "{Test Particles and Quasiperiodic Oscillations around Gravitational Aether Black Holes}",
      journal = {Galaxies},
         year = 2023,
        month = sep,
       volume = {11},
       number = {5},
          eid = {95},
        pages = {95},
          doi = {10.3390/galaxies11050095},
       adsurl = {https://ui.adsabs.harvard.edu/abs/2023Galax..11...95R}
}

@ARTICLE{2012MNRAS.426.1701B,
       author = {{Belloni}, T.~M. and {Sanna}, A. and {M{\'e}ndez}, M.},
        title = "{High-frequency quasi-periodic oscillations in black hole binaries}",
      journal = {\mnras},
         year = 2012,
        month = nov,
       volume = {426},
       number = {3},
        pages = {1701-1709},
          doi = {10.1111/j.1365-2966.2012.21634.x},
archivePrefix = {arXiv},
       eprint = {1207.2311},
 primaryClass = {astro-ph.HE},
       adsurl = {https://ui.adsabs.harvard.edu/abs/2012MNRAS.426.1701B}
}

@ARTICLE{2001A&A...374L..19A,
       author = {{Abramowicz}, M.~A. and {Klu{\'z}niak}, W.},
        title = "{A precise determination of black hole spin in GRO J1655-40}",
      journal = {\aap},
         year = 2001,
        month = aug,
       volume = {374},
        pages = {L19-L20},
          doi = {10.1051/0004-6361:20010791},
archivePrefix = {arXiv},
       eprint = {astro-ph/0105077},
 primaryClass = {astro-ph},
       adsurl = {https://ui.adsabs.harvard.edu/abs/2001A&A...374L..19A}
}

@ARTICLE{1998ApJ...499L..37M,
       author = {{Miller}, M. Coleman and {Lamb}, Frederick K.},
        title = "{Bounds on the Compactness of Neutron Stars from Brightness Oscillations during X-Ray Bursts}",
      journal = {\apjl},
         year = 1998,
        month = may,
       volume = {499},
       number = {1},
        pages = {L37-L40},
          doi = {10.1086/311335},
archivePrefix = {arXiv},
       eprint = {astro-ph/9711325},
 primaryClass = {astro-ph},
       adsurl = {https://ui.adsabs.harvard.edu/abs/1998ApJ...499L..37M}
}

@ARTICLE{1998ApJ...509L..37K,
       author = {{Klu{\'z}niak}, W.},
        title = "{General Relativistic Constraints on the Equation of State of Dense Matter Implied by Kilohertz Quasi-periodic Oscillations in Neutron-Star X-Ray Binaries}",
      journal = {\apjl},
         year = 1998,
        month = dec,
       volume = {509},
       number = {1},
        pages = {L37-L40},
          doi = {10.1086/311748},
archivePrefix = {arXiv},
       eprint = {astro-ph/9712243},
 primaryClass = {astro-ph},
       adsurl = {https://ui.adsabs.harvard.edu/abs/1998ApJ...509L..37K}
}

@ARTICLE{2022MPLA...3750064K,
       author = {{Khan}, Saeed Ullah and {Ren}, Jingli and {Rayimbaev}, Javlon},
        title = "{Circular motion around a regular rotating Hayward black hole}",
      journal = {Modern Physics Letters A},
         year = 2022,
        month = apr,
       volume = {37},
       number = {11},
          eid = {2250064},
        pages = {2250064},
          doi = {10.1142/S021773232250064X},
       adsurl = {https://ui.adsabs.harvard.edu/abs/2022MPLA...3750064K}
}

@ARTICLE{2023EPJC...83..506K,
       author = {{Kurbonov}, Nuriddin and {Rayimbaev}, Javlon and {Alloqulov}, Mirzabek and {Zahid}, Muhammad and {Abdulxamidov}, Farrux and {Abdujabbarov}, Ahmadjon and {Kurbanova}, Mukhabbat},
        title = "{Charged particles and Penrose process near charged black holes in Einstein-Maxwell-scalar theory}",
      journal = {European Physical Journal C},
         year = 2023,
        month = jun,
       volume = {83},
       number = {6},
          eid = {506},
        pages = {506},
          doi = {10.1140/epjc/s10052-023-11691-9},
       adsurl = {https://ui.adsabs.harvard.edu/abs/2023EPJC...83..506K}
}

@ARTICLE{1993ARA&A..31...93V,
       author = {{Verbunt}, Frank},
        title = "{Origin and evolution of X-ray binaries and binary radio pulsars.}",
      journal = {\araa},
         year = 1993,
        month = jan,
       volume = {31},
        pages = {93-127},
          doi = {10.1146/annurev.aa.31.090193.000521},
       adsurl = {https://ui.adsabs.harvard.edu/abs/1993ARA&A..31...93V}
}

@ARTICLE{2022EPJC...82.1110R,
       author = {{Rayimbaev}, Javlon and {Abdujabbarov}, Ahmadjon and {Abdulkhamidov}, Farukh and {Khamidov}, Vokhid and {Djumanov}, Sherzod and {Toshov}, Javohir and {Inoyatov}, Shukurillo},
        title = "{Quasiperiodic oscillation around charged black holes in Einstein-Maxwell-scalar theory}",
      journal = {European Physical Journal C},
         year = 2022,
        month = dec,
       volume = {82},
       number = {12},
          eid = {1110},
        pages = {1110},
          doi = {10.1140/epjc/s10052-022-11080-8},
       adsurl = {https://ui.adsabs.harvard.edu/abs/2022EPJC...82.1110R}
}

@ARTICLE{2022CQGra..39g5021R,
       author = {{Rayimbaev}, Javlon and {Bokhari}, Ashfaque Hussain and {Ahmedov}, Bobomurat},
        title = "{Quasiperiodic oscillations from noncommutative inspired black holes}",
      journal = {Classical and Quantum Gravity},
         year = 2022,
        month = apr,
       volume = {39},
       number = {7},
          eid = {075021},
        pages = {075021},
          doi = {10.1088/1361-6382/ac556a},
       adsurl = {https://ui.adsabs.harvard.edu/abs/2022CQGra..39g5021R}
}

@ARTICLE{2023EPJC...83..730Q,
       author = {{Qi}, Mai and {Rayimbaev}, Javlon and {Ahmedov}, Bobomurat},
        title = "{Charged particles and quasiperiodic oscillations around magnetized Schwarzschild black holes}",
      journal = {European Physical Journal C},
         year = 2023,
        month = aug,
       volume = {83},
       number = {8},
          eid = {730},
        pages = {730},
          doi = {10.1140/epjc/s10052-023-11912-1},
       adsurl = {https://ui.adsabs.harvard.edu/abs/2023EPJC...83..730Q}
}

@ARTICLE{Stuchlik2021JCAP...11..059S,
       author = {{Stuchl{\'\i}k}, Z. and {Vrba}, J.},
        title = "{Supermassive black holes surrounded by dark matter modeled as anisotropic fluid: epicyclic oscillations and their fitting to observed QPOs}",
      journal = {\jcap},
         year = 2021,
        month = nov,
       volume = {2021},
       number = {11},
          eid = {059},
        pages = {059},
          doi = {10.1088/1475-7516/2021/11/059},
archivePrefix = {arXiv},
       eprint = {2110.07411},
 primaryClass = {gr-qc},
       adsurl = {https://ui.adsabs.harvard.edu/abs/2021JCAP...11..059S}
}

@ARTICLE{2021JCAP...07..036F,
       author = {{Franzin}, Edgardo and {Liberati}, Stefano and {Mazza}, Jacopo and {Simpson}, Alex and {Visser}, Matt},
        title = "{Charged black-bounce spacetimes}",
      journal = {\jcap},
         year = 2021,
        month = jul,
       volume = {2021},
       number = {7},
          eid = {036},
        pages = {036},
          doi = {10.1088/1475-7516/2021/07/036},
archivePrefix = {arXiv},
       eprint = {2104.11376},
 primaryClass = {gr-qc},
       adsurl = {https://ui.adsabs.harvard.edu/abs/2021JCAP...07..036F}
}

\end{document}